\documentclass[a4paper]{aa} 

\usepackage[a4paper]{geometry}
\usepackage{geometry}
\usepackage{txfonts}
\usepackage[dvipsnames]{xcolor}

\usepackage{caption}
\usepackage{float}
\usepackage{dblfloatfix}

\usepackage{graphicx}

\usepackage{lscape}
\usepackage{amsmath}
\usepackage{subfig}
\usepackage{hyperref}
\usepackage[utf8]{inputenc}

\begin{document} 

\title{Planetary dynamos in evolving cold gas giants}

\authorrunning{Elias-López et al.}
\titlerunning{Cold Jupiters dynamos}
   \author{Albert Elias-López \inst{1,2,7},
          Fabio Del Sordo \inst{1,3,4},
          Daniele Viganò \inst{1,2,5},
          Clàudia Soriano-Guerrero \inst{1,2},
          Taner Akgün \inst{1,2},
          Alexis Reboul-Salze \inst{6},
          Matteo Cantiello \inst{7,8}
          }

\institute {Institut de Ci\`encies de I'Espai (ICE-CSIC), Campus UAB, Carrer de Can Magrans s/n, 08193 Cerdanyola del Vallès, Barcelona, Catalonia, Spain
\and
Institut d’Estudis Espacials de Catalunya (IEEC), 08860 Castelldefels, Barcelona, Catalonia, Spain
\and
INAF, Osservatorio Astrofisico di Catania, via Santa Sofia, 78 Catania, Italy
\and
Scuola Normale Superiore,
Piazza dei Cavalieri, 7
56126 Pisa, Italy
\and
Institute of Applied Computing \& Community Code (IAC3), University of the Balearic Islands, Palma, 07122, Spain
\and 
 Max Planck Institute for Gravitational Physics (Albert Einstein Institute), D-14476 Potsdam, Germany
\and
Center for Computational Astrophysics, Flatiron Institute, 162 5th Avenue, New York, NY 10010, USA
\and
Department of Astrophysical Sciences, Princeton University, Princeton, NJ 08544, USA
\\
\email{albert.elias@csic.es}
             }

   \date{Received ---------; accepted --------}

    \abstract
   {The discovery of thousands of exoplanets has started a new era of planetary science that expands our ability to characterize diverse planetary features. However, magnetic fields remain one of the least understood aspects of exoplanetary systems. A deeper understanding of planetary dynamos and the evolution of surface magnetic properties throughout the lifetime of a planet is a key scientific purpose. It has implications for planetary evolution, habitability, and atmospheric dynamics.
   }
   {We modeled the evolution of magnetic fields generated by dynamo action in cold gaseous giant planets. We explored the change in the morphology and strength of the magnetic field at different evolutionary stages, providing a comprehensive view of the planetary life-cycle.}
   {We solved the resistive magnetohydrodynamic (MHD) equations under the anelastic approximation with the 3D pseudo-spectral spherical shell MHD code MagIC. We employed 1D thermodynamical hydrostatic profiles taken from gas giant evolutionary models (using MESA) as the background states of our MHD models. The numerical integration led to saturated dynamo solutions. These calculations were performed with radial profiles corresponding to different planetary ages, so that we were able to interpret them as different snapshots of the planetary dynamo during the long-term planetary evolution.}
   {We characterized the magnetic field at different stages in the evolution of a cold gaseous planet. A transition from a multipolar to a dipolar-dominated dynamo regime occurs throughout the life of a Jovian planet. During the planetary evolution and the cooling-down phase, we observe a decrease in the average magnetic field strength near the dynamo surface as $\approx t^{-0.2}-t^{-0.3}$. This trend is compatible with previously proposed scaling laws. We also find that some dimensionless parameters evolve differently for the multipolar to the dipolar branch, possibly reflecting a force balance change.}
   {Our method captures the long-term evolution of the internal dynamo phases of magnetic fields by considering snapshots at different ages. We find a slow decay and a transition in the dynamo behavior. This approach can be extended to the study of hot gaseous planets, and it is a versatile method for predicting the magnetic properties of giant planets and for identifying promising candidates for exoplanetary low-frequency radio emission.}


\maketitle

\section{Introduction}\label{Sec: Introduction}

During the past decades, many numerical solutions to the magnetohydrodynamic (MHD) equations in convective spherical shells have been obtained under various assumptions and parameters. Simulations of this type produce an amplification and self-sustaining of magnetic fields that can recover some observed aspects related to a planetary dynamo in gas giants, such as the magnetic field morphology or the latitudinal variations in the atmospheric jets \citep[e.g.,]{Jones2011, Schubert&Soderlund2011}. High-resolution models are getting close to reproducing the internal dynamos of Earth and Jupiter (\cite{Schaefferetal2017, Gastine&Wicht2021}, respectively), including stochastic dipole reversals that resemble the geomagnetic field \citep{Glatzmaier&Coe2007, Medurietal2021}. In planetary science, MHD equations are typically expressed using dimensionless dynamo numbers: Rayleigh, Ekman, Prandtl, and magnetic Prandtl numbers, or combinations thereof. These numbers quantify the relative influence of various fluid forces such as the dissipative, buoyant, and Coriolis forces. There is an unavoidable computational caveat: The parameter space accessible through numerical simulations diverges significantly from the physical reality, and some parameters (Rayleigh and Ekman in particular) diverge by many orders of magnitude. The reason is that the range of the relevant spatial scales that are to be followed is too wide, that is, from the microscopical diffusion to the planetary scale for the global rotation or convection patterns.

To partially overcome this intrinsic drawback, some studies have used many numerical models to find scaling laws between the different dimensionless numbers characterizing the dynamo. For example, for rapidly rotating dipole-dominated dynamo solutions under the Boussinesq approximation, \cite{Christensen&Aubert2006} derived scaling laws that connect the dynamo parameters spanning at least two orders of magnitude. These relations should also arguably work in the real planetary regime, as the relative importance of each term in the Navier-Stokes equation (the force balance) is expected to be similar to numerical models \citep{Davidson2013, Yadavetal2016}. 

For the modeling of gas giants, the anelastic approximation \citep[e.g.,][]{Braginsky&Roberts1995, Gilman&Glatzmaier1981, Glatzmaier1984, Glatzmaier1985, Glatzmaier1985b, Lantz&Fan1999} is more appropriate than the Boussinesq approximation, as it allows for density variations but still effectively filters out the sound and magnetosonic waves. It relies on using a static, adiabatic, and spherically symmetric background reference state that is specified by density, gravity, temperature, and other thermodynamic variables. In addition, the velocity and magnetic fields are evolved together with the deviations from the background. These equations have been extensively used to model the magnetic field of gas giants and stars. Usually, the no-slip boundary conditions employed by geodynamo models are replaced by stress-free conditions because they reflect the nature of the outer atmosphere of the gas giants better. The simulations are more computationally demanding, and the numerical solutions have several features that are not seen in the Boussinesq approximation. For example, the development of Jupiter-like zonal flows, which promote weaker multipolar fields, and strong dipole fields, which can suppress these zonal flows via Lorentz forces, often compete in these models \citep[e.g.,][]{Grote&Busse2000, Simitev&Busse2003, Simitev&Busse2009, Sasakietal2011, Schrinneretal2012, Duarteetal2018}. Another example is the bistability found for not too large Rayleigh numbers: Dipolar-dominated and multipolar solutions coexist with identical parameters \citep{Gastineetal2012}, where different solutions can be reached by setting different initial conditions \citep{Schrinneretal2012}. In this context, \cite{Yadavetal2013} provided scaling laws for dynamo models under the anelastic approximation similar to those of \cite{Christensen&Aubert2006}. In this case, dipolar and multipolar solutions as well as a different range of density stratification and radial-dependent diffusivities were used. 

For a more realistic modeling of gas giants, the challenge is to incorporate the outer steep gradients of various thermodynamical profiles, since they imply very different timescales farther outward. Moreover, there is a steep outward decrease in the electrical conductivity because hydrogen is not in the metallic state in the outer planetary layers. This behavior was quantified along the Jovian adiabat by \cite{Frenchetal2012} and was used during the past decade in different dynamo simulations \citep{Gastine&Wicht2012, Jones2014, Gastineetal2014, Wichtetal2019}. These studies provided important results in terms of a comparison with the data provided by the ongoing Juno mission and earlier Jovian missions, and they highlighted the importance of incorporating a realistic background, although with the caveat on the nondimensional dynamo numbers mentioned above. In recent years, some works \citep{Gastine&Wicht2021, Yadavetal2022, Mooreetal2022} have added a stably stratified layer just below the region in which metallic hydrogen starts to mimic a helium-rain region due to hydrogen-helium demixing \citep{Klepeisetal1991, Nettelmann2015, Nettelmannetal2015}. This layer helps to naturally obtain alternating east-west zonal winds centered around the equator for the Jovian dynamo, as well as a highly axisymmetric magnetic field for the Saturn model.

By using the aforementioned scaling laws as well as observations, \cite{Christensenetal2009} determined that for both planets and fast-rotating stars, the energy flux determines the magnetic field strength. To match observational constraints, \cite{Reinersetal2009} expressed this law in terms of the mass $M$, luminosity $L$, and radius $R$ in a more simplified form. Using this scaling law and the analytical evolutionary tracks for substellar objects in \citep{Burrows&Liebert1993, Burrowsetal2001}, \cite{Reiners&Christensen2010} provided a scenario for the evolution of the magnetic field, with which they obtained a steadily weakening (a factor of $\approx$10 over around 10 Gyr) of the magnetic field at the dynamo surface.

We aim to address the long-term evolution of the dynamo action in Jupiter-like planets through an alternative approach. We perform 3D anelastic dynamo simulations with a background corresponding to different ages of the long-term planetary evolution. By comparing how the solutions change from one age to the next and recalling the intrinsic caveats related to the accessible ranges of nondimensional numbers, we evaluate the trend in morphology and intensity changes that an internal dynamo in cold gas giants can undergo during its evolution over billions of years.

This work is organized as follows: The overall method and the internal thermodynamical profiles from the evolutionary code MESA are described in Sect. \ref{Sec: Method}, where we also summarize our 3D dynamo models that were performed with the MagIC code. In Sect. \ref{Sec: Results} we show the main results of our parameter exploration, interpret simulations representative of different evolutionary stages, and compare them to results from other works. We finally conclude in Sect. \ref{Sec: Conclusions}.

\section{Method}\label{Sec: Method}

It is currently not possible to simulate the realistic time evolution of the radial dependence of the thermodynamic quantities of a 3D planetary dynamo. The main reason is that the timescale associated with the convection inside gas giants tends to be on the order of years or decades. In contrast, the secular planetary cooling and contraction are appreciable at timescales on the order of gigayears. Due to this timescale separation of at least six orders of magnitude, a set of fixed backgrounds can be considered, and the dynamo models can be evolved for each of them. In other words, we aim at having different snapshots, that is, 3D dynamo solutions, each with fixed radial thermodynamical profiles that correspond to a given age of the 1D long-term evolutionary models. To schematically summarize, we employed a method that followed these steps:
\begin{enumerate}
    \item Evolve a standard evolutionary 1D model of a contracting, nonirradiated gas giant over 10 Gyr.
    \item For a given age of the evolutionary model, implement the radially dependent thermodynamical profiles as the background state of an anelastic spherical shell MHD model, with a given choice of the nondimensional numbers.
    \item Evolve the 3D MHD equations in a spherical shell domain under the anelastic approximation and reach a self-sustained dynamo solution. Then let it evolve long enough (typically up to some ten thousand years of physical timescales) to average out the typical fluctuations and ensure statistical significance.
    \item Repeat the process from step 2, where some of the nondimensional dynamo numbers rescale (compared to the first simulation) according to the relative variation of the involved thermodynamic quantities. Note that all the nondimensional numbers can be rescaled in this way, so that further assumptions about the rotation rate and viscosity, for instance, are required, as discussed below.
\end{enumerate}
The process can then be repeated for another planetary model or a different choice of the reference values of the nondimensional numbers. In this approach, the cooling information is included in the trend of the dynamo numbers in the simulation sequence.

\subsection{Internal structure}

\subsubsection{Long-term evolution}

To model the evolutionary change of radially dependent thermodynamic quantities of gas giants, we used the public code MESA\footnote{\url{https://github.com/MESAHub/mesa}}  \citep{Paxtonetal2011, Paxtonetal2013, Paxtonetal2015, Paxtonetal2018, Paxtonetal2019}. This is a one-dimensional code that solves the time-dependent stellar structure equations and is capable of evolving low mass bodies, including brown dwarfs and gas giants \citep[see][]{Paxtonetal2013}. The equations we solved are the conservation of mass, hydrostatic equilibrium, energy conservation, and the energy transport equation, respectively,
\begin{gather}
    \frac{dm}{dr} = 4 \pi r^2 \rho ~,
    \label{Eq: MESA mass conservation}\\
    \frac{dP}{dm} = - \frac{G m}{4\pi r^4} ~,
    \label{Eq: MESA hydrostatic equilibrium}\\
    \frac{dL}{dm} = - T \frac{ds}{dt} ~,
    \label{Eq: MESA luminosity}\\
    \frac{dT}{dm} = - \frac{G m T}{4\pi r^4 P} \nabla ~,
    \label{Eq: MESA energy transport}
\end{gather}
where $m$ is the mass enclosed within a radius $r$, $\rho$ is the density, $P$ is the pressure, $G$ is the gravitational constant, $s$ is the specific entropy, $T$ is the temperature, $L$ is the internal luminosity, and $\nabla \equiv d \ln T / d \ln P$ is the logarithmic temperature gradient, which was set to the smallest between the adiabatic gradient and the radiative gradient. In the energy equation~\ref{Eq: MESA luminosity}, the only source term we considered is the gravitational contraction. We neglected additional sources such as stellar irradiation \citep{Guillotetal96} or internal heat deposition \citep{Komacek&Youdin2017, Thorngren&Fortney2018} from tidal \citep{Bodenheimeretal01} or Ohmic dissipation (e.g., \citealt{Batygin&Stevenson2010, Pernaelat2010}), or chemical processes such as hydrogen dissociation and recombination \citep{Tanetal19}. These additional terms are fundamental for hot Jupiters (see \citealt{Fortneyetal2021} for a review), but negligible for cold, weakly irradiated planets. The set of equations is closed using the MESA equation of state \citep{Paxtonetal2019}, which for the gas giant ranges of interest, is substantially the interpolation of the Saumon-Chabrier-van Horn equation of state for H-He mixtures \citep{Saumonetal1995}.

\begin{figure}[t!]
\centerline{\includegraphics[width=.92\hsize]{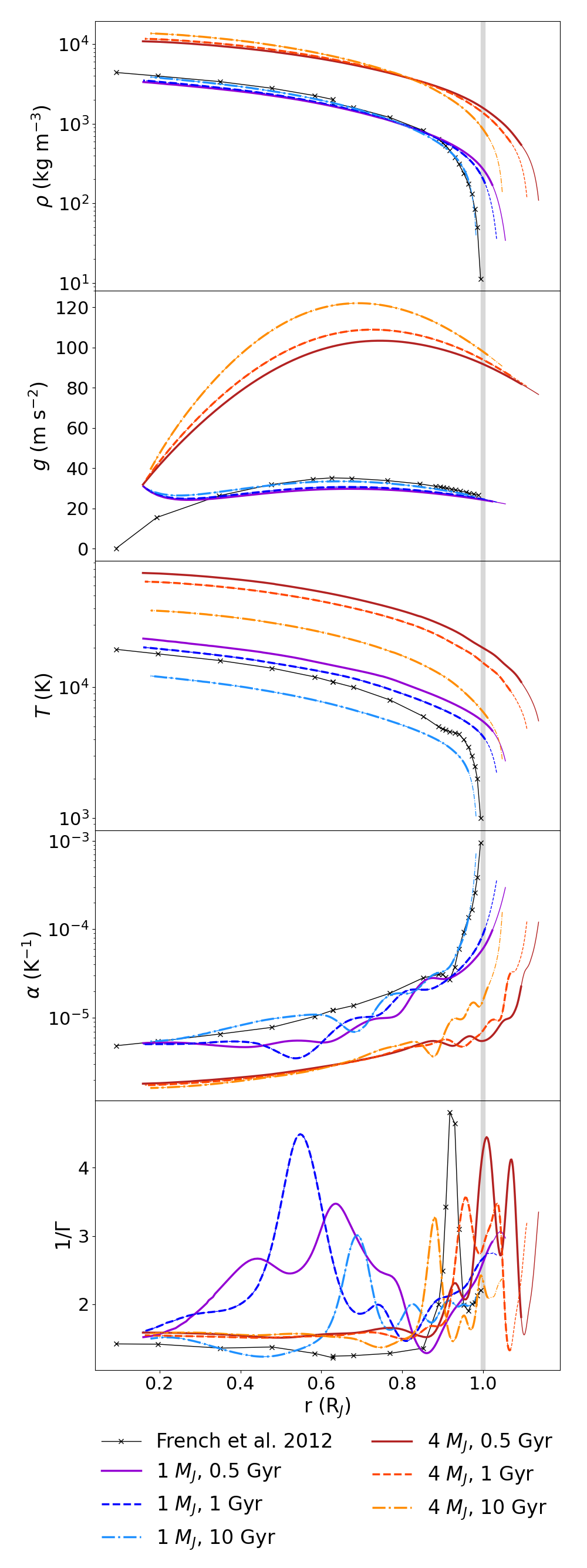}}
\caption{MESA hydrostatic profiles of 1 and 4 $M_J$ at different evolutionary times, cut at an outer density $\approx 100$ times (thin extended lines) or $\approx 20$ times (thick lines only) that is lower than the innermost radius of the isentropic shell, just outside the core-envelope boundary. The gray lines show the Jovian values according to the popular \cite{Frenchetal2012} model, and the vertical gray band reflects the current Jovian radius as a reference. From top to bottom: density $\rho(r)$, temperature $T(r)$, gravity $g(r)$, thermal expansion coefficient $\alpha(r)$, and the inverse of the Gr\"uneisen parameter $\Gamma(r)$.}
\label{Fig: CJ Profiles}
\end{figure}

These 1D evolutionary models do not incorporate either diluted cores or hydrogen-helium demixing layers, that is, possible stratified layers in the convection interior, of the type mentioned in Sect. \ref{Sec: Introduction}. For all our models, we assumed an interior inert rocky core of 10 $M_{\oplus}$ with a homogeneous density $\rho_c=10$~g~cm$^{-3}$, and a fixed solar composition for the envelope. To illustrate the evolutionary changes, we show in Fig. \ref{Fig: CJ Profiles} the different profiles for two different planets with masses 1 and 4 $M_J$ at ages 0.5, 1, and 10 Gyr. For comparison, we also show the profiles from the widely employed results of \cite{Frenchetal2012} for the interior of Jupiter. As reported by \citep{Paxtonetal2013}, during the evolution of the planet, the radius shrinks slowly, and after a few million years of evolution, this is independent of the chosen initial planetary radius. The planet slowly shrinks, and at early ages, the total radius of the $1\, M_J$ model is therefore larger than that of the current Jupiter. The internal structure is always characterized by a very thin radiative layer with a thick, fully convective isentropic shell that encloses the inert core. The higher planetary mass ($4\, M_J$) shows a larger radius, higher gravity and temperature, and lower thermal expansion coefficients $\alpha$. However, the trends of the profile with age are similar to the $1\, M_J$ case. Moreover, all models exhibit a nontrivial oscillating behavior of the Gr\"uneisen parameter $\Gamma$ (bottom panel; see below for the definition).

\subsubsection{Background-state implementation}
\label{Sec: Background state implementation}

To run an anelastic MHD model, a series of thermodynamical quantities is required to be expressed as a function of radius. We implemented the MESA $\rho(r)$, $T(r)$, $g(r)$, thermal expansion coefficient $\alpha(r)$, and the Grüneisen parameter $\Gamma(r)$, all shown in Fig. \ref{Fig: CJ Profiles}, into a 3D model. We obtained $\alpha$ and $\Gamma$ in terms of the readily available MESA thermodynamic profiles
\begin{equation}
    \Gamma = \left(\frac{\partial ln T}{\partial ln \rho}\right)_s - 1 \equiv \Gamma_3 - 1 ~, \quad \alpha = - \frac{1}{\rho} \left( \frac{\partial \rho}{\partial T} \right)_P = - \frac{\chi_T}{T \chi_\rho} ~,
\end{equation}
where
\begin{equation*}
    \chi_\rho \equiv \left(\frac{\partial ln P}{\partial ln \rho}\right)_T ~, \quad \chi_T \equiv \left(\frac{\partial ln P}{\partial ln T}\right)_\rho ~.
\end{equation*}
Transport coefficients, which we discuss and prescribe below, are the other possible thermodynamical quantities that can show a radial dependence.

To define the radial domain for the 3D models, we cut the MESA profiles both at the inner and outer radial domain. The hydrostatic background is mostly isentropic due to the efficient convection. Close to the solid inert core, the profiles show a decrease in entropy, leading to a shallow stratified layer that is due to the boundary conditions. Since a more realistic modeling of the core is beyond the purpose of this study, we considered as the inner boundary of the convective shell for each model the region in which the profile is isentropic and cut out the innermost $\approx 2$-$3\%$ (in radius) of the shell case by case. We do not expect these inner cuts to lead to significant differences because the results of \cite{Mooreetal2022} showed that replacing a diluted core with a solid compact core of the same radius has little effect on the strength and morphology of the dynamo.

The density profiles usually spanned more than six orders of magnitude (the typical outermost layer in MESA is at a fraction of a bar), with the largest drop in the 1\% outermost part of the planet, where no dynamo is expected. Spherical shell dynamo models cannot handle too large density contrasts. Therefore, as is common in other anelastic dynamo models, we cut the external layers to reduce the density ratio. We ensured that we always kept the pressure at which hydrogen metallization starts ($\approx$1 Mbar) inside the domain. The maximum ratio we considered is $\rho_i/\rho_o \approx 100$, and most of our models took values of $ \rho_i/\rho_o \approx 20$, which corresponds to approximate values for the number of density scale heights, $N_\rho:= ln(\rho_i/\rho_o)$, of 4.6 and 3.0, respectively. The thick profiles shown in Fig. \ref{Fig: CJ Profiles} have $N_\rho \approx$ 3, and the thin endings represent an extension up to $N_\rho \approx$ 4.6. Throughout this work, we use the subindex notation $o$ ($i$) for the value at the outer (inner) shell. These cuts define $r_o$, $\rho_o$, $T_o$, and $P_o$ as well as their inner values, which are listed in Table \ref{Tab: Cold Jupiter input parameters}. The thin radiative outer layer usually ends at about $\approx$ 1 bar, well above the external cut. Finally, all profiles except for $\Gamma(r)$ were normalized to make the outer values equal to unity. This is a common practice for many 3D hydrodynamic codes as the fundamental units are not the physical units. MagIC, for example, works with units of the shell thickness as the length scale and with viscous timescales for units of time (see Sect. \ref{Sec: 3D numerical model MagIC} for the exact details).

After the profiles were cut and normalized to the inner or outer boundary (as required by the code implementation; see below), we used high-degree polynomials to generally fit any shape the profiles can take. For $\rho(r)$, $T(r)$, and $g(r)$, we employed a 20-degree Taylor series, which was enough to smoothly fit all cases tested here. $\alpha(r)$ and $\Gamma(r)$ have some peaks and valleys that complicate the fitting procedure, however. With polynomials with degrees below $\approx$ 100-120, we found that the wiggles were poorly fit. We adopted a 150-degree polynomial for both,
\begin{equation}
\begin{aligned}
    \Big( \rho(r),T(r),g(r) \Big) = \sum_{n=0}^{20} \Big( \rho_n, T_n, g_n \Big) r^n \,,    \\
    \Big( \alpha(r), \Gamma(r) \Big) = \sum_{n=0}^{150} \Big( \alpha_n, \Gamma_n \Big) r^n\, , 
\end{aligned}
\end{equation}
where $r$ ranges from $r_i$ to $r_o$. A change in the reference point of the expansion did not quantitatively improve the fits (we also tried with powers of $(r - r_o)$, $(r - r_i)$ and $(r - r_o/2)$). Therefore, for simplicity, we opted for powers of $r$, that is, a MacLaurin series. We also verified that the radial resolution we employed in the MHD code, that is, the number of Chebyshev polynomials (see Sect. \ref{Sec: 3D numerical model MagIC}), was more than enough to satisfactorily reproduce even the most complex radial profiles given by the implemented MacLaurin expansion.

Unlike the fit model used by \cite{Jones2014}, the background profiles taken from MESA are almost but not exactly isentropic. To quantify this deviation, we considered the quantity $|ds/dr|\cdot r/s$, which usually takes values of $10^{-4} - 10^{-5}$ with maxima near the outer cut regions of $10^{-3}$. This might lead to a slight energy imbalance resulting from the radial background profile itself. To ensure that this did not influence the overall dynamics, we analyzed the energy balances, that is, we compared the buoyancy power with viscous and Ohmic dissipation (see Sec \ref{Sec: Diagnostic parameters} for more details).

\subsubsection{Transport coefficients}
\label{Sec: Transport coefficients}

\begin{figure}[t!]
\centerline{\includegraphics[width=\hsize]{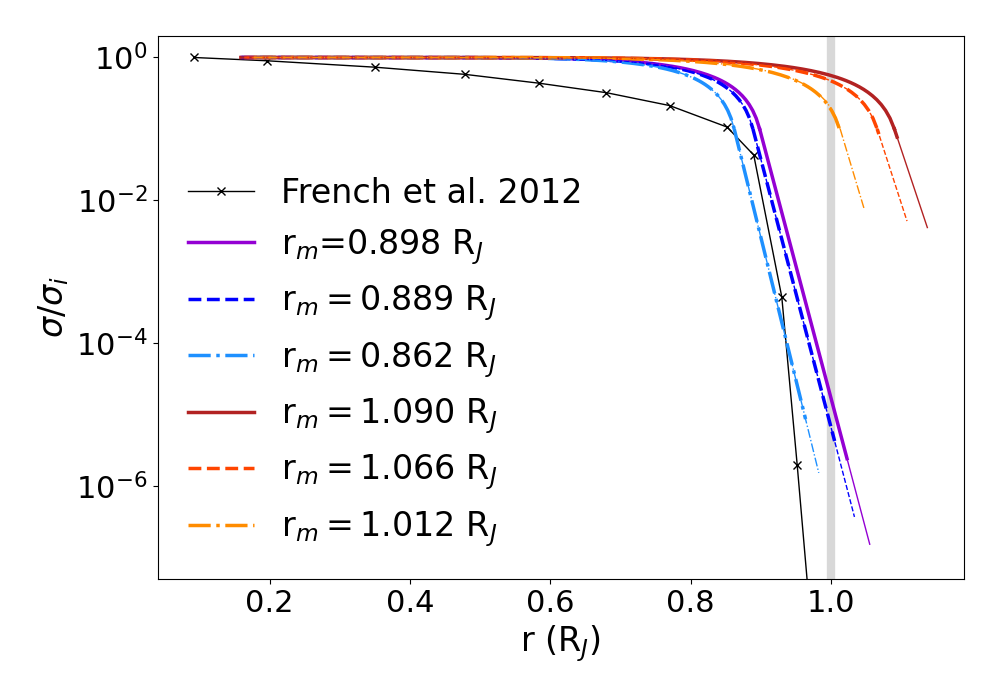}}
\caption{Electrical conductivities normalized to their inner values ($\sigma/\sigma_i$) obtained from eq.~(\ref{Eq: Analytical conductivity Gom10}) for the same models as shown in Fig. \ref{Fig: CJ Profiles}. The values of $r_m$ in the legend correspond to the start of the exponential decay. The \cite{Frenchetal2012} profile was also normalized.}
\label{Fig: CJ profiles conductivities}
\end{figure}

The profiles of transport coefficients did not come directly from MESA. Although some important ingredients have evolved, such as the particle density, realistic profiles for diffusivities require proper ab initio calculations. In particular, the electrical and thermal conductivities have to take into account the degenerate state of the electron population. The pressure ionization (rather than thermal) is non-negligible in the dense but relatively cold (compared to stars) convective interior. Moreover, the dynamo region arguably corresponds to the transition to the metallic phase for hydrogen. In this sense, \cite{Frenchetal2012} calculated the electric conductivity $\sigma$ for a set of $(T,\rho)$ pairs along the modeled Jupiter adiabat. Further models about transport coefficients in pure H-He mixtures have been developed (usually neglecting the contribution of thermally ionized alkali metals, which becomes relevant where hydrogen is molecular; \citealt{Kumaretal2021}), with relative differences in the values of $\sigma$ by factors of a few (see \citealt{Bonitzetal2024} for a recent review). In any case, the trend is that there is a continuous and steep increase in the conductivity for increasing pressure up to around 1 Mbar, after which the dependence on both temperature and pressure (or density) is much milder.

To capture these fundamental properties, we adopted the electrical conductivity profile first defined in \cite{GomezPerezetal2010}, which consists of an approximately constant conductivity in the innermost hydrogen metallic region, with a polynomial plus exponential decay toward the outer molecular region,\footnote{A spurious contribution to Ohmic dissipation is produced very close to the outer boundary. We checked how relevant it is in terms of affecting the saturated solution by comparing representative runs with what is obtained by setting a thin perfectly insulating fluid layer in the outermost $\approx$1-2$\%$ (in radius) of the domain, by using the parameter {\tt $r\_{LCR}$} in MESA. In all cases, we found negligible differences in all relevant time-averaged diagnostics (Sect. \ref{Sec: Diagnostic parameters}) within the stochastic oscillations of the solutions.}
\begin{equation}
    \frac{1}{{\tilde\lambda}(r)} := \tilde{\sigma}(r) = \begin{cases} 1+ (\sigma_m-1)\left(\dfrac{r-r_i}{r_m-r_i}\right)^{-a} \quad  r<r_m ~, \\
    \sigma_m e^{-a \left(\frac{r-r_m}{r_m-r_i}\right)\frac{\sigma_m-1}{\sigma_m}}
    \hspace{1.7cm} r\geq r_m ~,\end{cases},
\label{Eq: Analytical conductivity Gom10}
\end{equation}
where $\tilde{\sigma}$ and $\tilde{\lambda}$ are the normalized conductivity and magnetic diffusivity, respectively. The actual physical relation, $\lambda=(\mu_0\sigma)^{-1}$, with $\mu_0$ being the vacuum magnetic permeability, is simplified when the quantities are normalized to their innermost values: $\tilde{\sigma}=\sigma/\sigma(r_i)$ and $\tilde{\lambda}=\lambda/\lambda(r_i)$. This expression ensures that both $\tilde \lambda$ and $d{\tilde \lambda}/dr$ are continuous at $r_m$, and that they qualitatively reproduce the main features. Moreover, it allowed us to compare results with several previous works that have employed this profile for gas giant convection and dynamo modeling \citep{Duarteetal2013, Duarteetal2018, Wichtetal2019, Wichtetal2019b, Gastine&Wicht2021}. They used values of $\sigma_m$ and $a$ that ranged from approximately 0.9 to 0.01 and from 1 to 25. In our models, we fixed $\sigma_m$ = 0.1 and $a$ = 7, while $r_m$ was chosen as the radius at which each MESA planetary profile reached 1 Mbar, that is, the approximate pressure above which hydrogen is thought to undergo metallization. The profiles of $\tilde{\sigma}$ are shown in Fig. \ref{Fig: CJ profiles conductivities} for the same representative models as in Fig. \ref{Fig: CJ Profiles}.

On the other hand, for simplicity, we kept the kinematic viscosity $\nu$ and thermal diffusivity $\kappa$ constant within the same model. In Sect. \ref{Sec: Dependence on Prandt number}, we return to the impact and caveats of this choice and the related assumption about the Prandtl numbers (Sect. \ref{Sec: Parameter evolution}).

\subsection{3D numerical dynamo model}
\label{Sec: 3D numerical model MagIC}

The next step is fundamental: We performed 3D MHD spherical shell simulations using the public code MagIC\footnote{\url{https://github.com/magic-sph/magic}. We used the version 6.3: \url{https://zenodo.org/records/3564626}} with the anelastic approximation \citep{Gastine&Wicht2012}. This is a pseudo-spectral code that uses the spherical harmonic decomposition in the angular directions, that is, $\theta$ and $\phi$, and Chebyshev polynomials in the radial direction $r$. MagIC has been used for both stellar and planetary models, including for Jupiter and Saturn dynamos, for instance, \cite{Duarteetal2018, Wichtetal2019, Gastine&Wicht2021, Yadavetal2022}, and it was tested in convection and dynamo benchmarks \cite{Christensenetal2001, Jonesetal2011}. We employed the anelastic approximation, which is typically used for modeling the density-stratified low-Mach number convection flows in gas giants and stars \cite{Braginsky&Roberts1995, Lantz&Fan1999}.

The shell is filled with a finitely conducting fluid rotating along the vertical axis $\boldsymbol{\hat{z}}$ with a constant angular velocity $\Omega$, and the background profiles, which we set up as explained in Sect. \ref{Sec: Background state implementation}. The geometry is set by the aspect ratio $\eta:= r_i/r_o$. We work in dimensionless units using the shell thickness $d:= (r_o - r_i)$ as the length unit and the viscous diffusion timescale $d^2/\nu$ as the time unit. The magnetic fields are in units of $(\rho_o \mu_o \lambda_i \Omega)^{1/2}$, where $\lambda_i=1/(\mu_0\sigma(r_i))$ is the magnetic diffusivity at the inner boundary. The convection is set by a fixed entropy gradient $\Delta s$, and this difference serves as the nondimensional units for $s$. 

The equations we solved were the mass continuity equation, the momentum equation, the entropy equation, the induction equation, and the solenoidal constraint for the magnetic field,
\begin{equation}
\mathbf{\nabla}\cdot(\tilde{\rho}\mathbf{u})=0 ~,
\label{Eq: Magic mass continuity}
\end{equation}

\begin{equation}
\begin{aligned}
	\dfrac{\partial \mathbf{u}}{\partial t}+\mathbf{u}\cdot\mathbf{\nabla}\mathbf{u}
	= -\mathbf{\nabla}\left({\dfrac{p'}{\tilde{\rho}}}\right) - \dfrac{2}{E}\mathbf{e_z}\times\mathbf{u}
	- \dfrac{Ra}{Pr}\tilde{g}\tilde{\alpha}\tilde{T} \,s'\,\mathbf{e_r} + \\ 
	+\dfrac{1}{Pm\,E \,\tilde{\rho}}\left(\mathbf{\nabla}\times \mathbf{B}
	\right)\times \mathbf{B}+ \dfrac{1}{\tilde{\rho}} \mathbf{\nabla} \cdot \mathbf{S} ~,
\end{aligned}
\label{Eq: Magic momentum evolution}
\end{equation}

\begin{equation}
\begin{aligned}
	\tilde{\rho}\tilde{T}\left(\dfrac{\partial s'}{\partial t} +
	\mathbf{u}\cdot\mathbf{\nabla} s'\right) =
	\dfrac{1}{Pr}\mathbf{\nabla}\cdot\left(\tilde{\rho}\tilde{T}\mathbf{\nabla} s'\right) +
	\dfrac{Pr\,Di}{Ra} (Q_\nu +  Q_\lambda) ~,  
\end{aligned}
\label{Eq: Magic entropy evolution}
\end{equation}

\begin{equation}
	\dfrac{\partial \mathbf{B}}{\partial t} = \boldsymbol{\nabla} \times \left( \mathbf{u}\times\mathbf{B}\right)-\dfrac{1}{Pm_i}\boldsymbol{\nabla}\times\left(\tilde{\lambda}\boldsymbol{\nabla}\times\mathbf{B}\right) ~,
\label{Eq: Magic induction}
\end{equation}

\begin{equation}
	  \mathbf{\nabla}\cdot\mathbf{B}=0 ~,
\label{Eq: Magic solenoidal condition}
\end{equation}
where the traceless rate-of-strain tension $S_{ij}$ an the viscous and Ohmic heating terms, $Q_\nu$ and $Q_\lambda$ are defined by
\begin{equation*}
\begin{aligned}
    S_{ij} \equiv 2 \tilde{\rho} \left( e_{ij} - \frac{1}{3}\delta_{ij} \boldsymbol{\nabla}\cdot \mathbf{u} \right) ~, \quad 
    e_{ij} \equiv \frac{1}{2} \left( \frac{\partial u_i}{\partial x_j} + \frac{\partial u_j}{\partial x_i} \right) ~, \\
    Q_\nu \equiv 2 \tilde{\rho} \left( e_{ij} e_{ij} - \frac{1}{3} (\boldsymbol{\nabla} \cdot \mathbf{u})^2 \right) ~, \quad Q_\lambda \equiv \dfrac{\tilde{\lambda}}{Pm^2\,E}\left(\mathbf{\nabla}
	\times\mathbf{B}\right)^2 ~.
\end{aligned}
\end{equation*}
The dimensionless Ekman, Rayleigh, Prandtl, and magnetic Prandtl numbers are respectively defined as 
\begin{align*}
    E \equiv \frac{\nu}{\Omega d^2}~, \quad 
    Ra \equiv \frac{\alpha_o g_o T_o d^3 \Delta s}{c_p \nu \kappa}~, \quad
    Pr \equiv \frac{\nu}{\kappa}~, \quad 
    Pm \equiv \frac{\nu}{\lambda_i}~.
\end{align*}
All the quantities marked with a tilde (Sect. \ref{Sec: Background state implementation} and \ref{Sec: Transport coefficients}) are static in time, radially dependent, and normalized to their outer values, except for $1/\lambda$, which because of its fast decay in the outer radial regions, was normalized to its innermost value for practical reasons.

\subsubsection{Boundary conditions}

We assumed stress-free and impenetrable boundary conditions for the velocity at the inner and outer radii, $r = r_i, r_o$,
\begin{equation*}
    u_r = \frac{\partial}{\partial r} \left( \frac{u_\theta}{r} \right) = \frac{\partial}{\partial r} \left( \frac{u_\phi}{r} \right) = 0 ~.
\end{equation*}
When stress-free boundary conditions are used at both boundaries, MagIC gives the option to ensure angular momentum conservation, which we have used. We employed constant entropy at both boundary conditions, 
\begin{equation*}
    s'(r=r_0)=0 ~, \quad s'(r=r_i)=1 ~.
\end{equation*}
The material outside the outer radius was electrically insulating, that is, the magnetic field matched a potential field. At the inner boundaries, we imposed a perfectly conducting core.

\subsubsection{Numerical technique}

MagIC solves the set of equations (\ref{Eq: Magic mass continuity} - \ref{Eq: Magic solenoidal condition}) with the above boundary conditions by expanding the mass flux and the magnetic fields into poloidal and toroidal potentials, 
\begin{equation*}
\begin{gathered}
    \tilde{\rho} \mathbf{u} = \boldsymbol{\nabla}\times(\boldsymbol{\nabla}\times W\,\mathbf{e_r}) +
    \boldsymbol{\nabla}\times Z\,\mathbf{e_r} ~, \\
    \mathbf{B} = \boldsymbol{\nabla}\times(\boldsymbol{\nabla}\times g\,\mathbf{e_r}) +
    \boldsymbol{\nabla}\times h\,\mathbf{e_r} ~.
\end{gathered}
\label{Eq: Magic poloidal/toroidal decomposition}
\end{equation*} 
The quantities $W$, $Z$, $g$, $h$, $s'$ and $p'$ are expanded up to $l_{max}$ in spherical harmonic degree and $N_C$ in Chebyshev polynomials, where we always set $N_C=N_r-1$. The equations are time-stepped by advancing nonlinear and Coriolis terms using an explicit second-order Adams-Bashforth scheme, and the remaining terms were time-advanced using the implicit Crank-Nicolson algorithm (for more details, see \cite{Glatzmaier1984, Christensen&Wicht2007}).

\begin{table*}[ht]
\centering
\tiny
\caption{Parameters of the 1D MESA models for the 3D MagIC input.}
\begin{tabular}{@{}ccccccccccccccccccccc@{}}
\hline \hline \\[-2.0ex]
  Model         & $N_\rho$ & $\rho_o$(g$\cdot$cm$^{-3}$) & $\Delta T$ (K) & $T_o$ (K) & $P_o$ (kbar) & $r_o$(R$_J$) &  $\eta$ & $M$ & $\chi_m$ & $E$   & $Ra$    \\ \hline  \\[-2.0ex]
  1$M_J$ 0.4 Gyr& 2.99   & 0.165   & 19782  & 4858   & 55.6 & 1.030  & 0.164  & 47.36 & 0.874 & 1.08$\cdot$10$^{-5}$ & 1.27$\cdot$10$^{9}$ \\
  1$M_J$ 0.5 Gyr& 2.98   & 0.171   & 18847  & 4665   & 57.3 & 1.023  & 0.156  & 45.58 & 0.878 & 1.08$\cdot$10$^{-5}$ & 1.22$\cdot$10$^{9}$ \\
  1$M_J$ 0.7 Gyr& 2.99   & 0.172   & 17345  & 4306   & 54.8 & 1.014  & 0.166  & 48.03 & 0.881 & 1.12$\cdot$10$^{-5}$ & 1.05$\cdot$10$^{9}$ \\
  1$M_J$ 1  Gyr & 2.99   & 0.177   & 16158  & 4002   & 54.5 & 1.005  & 0.159  & 46.91 & 0.884 & 1.12$\cdot$10$^{-5}$ & 9.81$\cdot$10$^{8}$ \\
  1$M_J$ 1.5 Gyr& 2.98   & 0.181   & 15004  & 3716   & 54.0 & 0.998  & 0.1628 & 47.47 & 0.884 & 1.15$\cdot$10$^{-5}$ & 8.78$\cdot$10$^{8}$ \\
  1$M_J$ 2.1 Gyr& 2.99   & 0.181   & 13999  & 3428   & 51.0 & 0.992  & 0.169  & 49.61 & 0.887 & 1.18$\cdot$10$^{-5}$ & 7.87$\cdot$10$^{8}$ \\
  1$M_J$ 2.8 Gyr& 2.99   & 0.184   & 13278  & 3220   & 50.0 & 0.987  & 0.170  & 49.77 & 0.890 & 1.20$\cdot$10$^{-5}$ & 7.33$\cdot$10$^{8}$ \\
  1$M_J$ 3.5 Gyr& 2.99   & 0.187   & 12424  & 2983   & 49.2 & 0.981  & 0.171  & 50.07 & 0.889 & 1.21$\cdot$10$^{-5}$ & 6.71$\cdot$10$^{8}$ \\
  1$M_J$ 5  Gyr & 2.99   & 0.190   & 11742  & 2779   & 48.3 & 0.975  & 0.172  & 52.29 & 0.892 & 1.23$\cdot$10$^{-5}$ & 6.22$\cdot$10$^{8}$ \\
  1$M_J$ 6.5 Gyr& 2.98   & 0.192   & 11031  & 2591   & 47.3 & 0.970  & 0.183  & 53.35 & 0.892 & 1.28$\cdot$10$^{-5}$ & 5.52$\cdot$10$^{8}$ \\
  1$M_J$ 10 Gyr & 3.00   & 0.193   & 10180  & 2327   & 44.7 & 0.964  & 0.184  & 53.35 & 0.894 & 1.30$\cdot$10$^{-5}$ & 4.98$\cdot$10$^{8}$ \\ \hline\\ [-2.0ex]
  1$M_J$ 1 Gyr  & 2.30   & 0.351   & 14973  & 5220   & 240  & 0.967  & 0.165  & 26.48 & 0.919 & 1.23$\cdot$10$^{-5}$ & 7.92$\cdot$10$^{8}$ \\ 
  1$M_J$ 1 Gyr  & 3.68   & 0.0881  & 17093  & 3100   & 14.9 & 1.022  & 0.156  & 89.66 & 0.869 & 1.08$\cdot$10$^{-5}$ & 1.10$\cdot$10$^{9}$ \\
  1$M_J$ 1 Gyr  & 4.58   & 0.0358  & 17962  & 2231   & 3.46 & 1.034  & 0.155  & 214.0 & 0.860 & 1.05$\cdot$10$^{-5}$ & 1.20$\cdot$10$^{9}$ \\  \hline \\[-2.0ex]
  0.3$M_J$ 5 Gyr& 1.10   & 0.545   & 3022   & 3754   & 563  & 0.650  & 0.275  & 16.89 & 0.879 & 3.61$\cdot$10$^{-5}$ & 3.18$\cdot$10$^{7}$  \\
  0.7$M_J$ 5 Gyr& 2.98   & 0.144   & 8927   & 2243   & 25.1 & 0.944  & 0.190  & 53.69 & 0.843 & 1.37$\cdot$10$^{-5}$ & 4.01$\cdot$10$^{8}$  \\
  2$M_J$ 5 Gyr  & 2.98   & 0.338   & 20486  & 4195   & 198  & 1.015  & 0.167  & 50.13 & 0.955 & 1.12$\cdot$10$^{-5}$ & 1.24$\cdot$10$^{9}$  \\
  4$M_J$ 5 Gyr  & 2.98   & 0.671   & 38765  & 6762   & 1060 & 1.025  & 0.159  & 47.94 & 1.000 & 1.08$\cdot$10$^{-5}$ & 2.49$\cdot$10$^{9}$  \\ \hline \\[-2.0ex]
  4$M_J$ 0.5 Gyr& 4.60   & 0.109   & 69086  & 5533   & 33.6 & 1.137  & 0.140  & 32.20 & 0.959 & 8.37$\cdot$10$^{-6}$ & 6.50$\cdot$10$^{9}$  \\ 
  4$M_J$ 1 Gyr  & 4.58   & 0.120   & 59273  & 4829   & 33.9 & 1.109  & 0.148  & 31.26 & 0.961 & 8.98$\cdot$10$^{-5}$ & 5.02$\cdot$10$^{9}$  \\ 
  4$M_J$ 10 Gyr & 4.60   & 0.137   & 36481  & 2880   & 27.7 & 1.048  & 0.169  & 31.81 & 0.967 & 1.06$\cdot$10$^{-5}$ & 2.42$\cdot$10$^{9}$  \\ \hline \hline \\ [-2.0ex]
\end{tabular}
\label{Tab: Cold Jupiter input parameters}
\tablefoot{Each line describes a different planetary model at a specific time and density cut, with the corresponding values of $\eta=r_i/r_o$, $\chi_m = r_m/r_o$, the dimensionless mass of the shell $M=4\pi\int_{r_i}^{r_o} r^2 \frac{\rho(r)}{\rho_o} dr$, and the dynamo parameters $E$, $Ra$ (after calibrating them in one case, as described in the text).}
\end{table*}

\subsubsection{Diagnostic parameters}
\label{Sec: Diagnostic parameters}

To characterize the numerical dynamo solutions, we made use of several diagnostic quantities. We typically took averages in time or in space, either over the whole volume $V$ or over spherical surfaces, to show the radial dependence, 
\begin{equation*}
\begin{gathered}
    ||a|| \, (r,t) = \int a(r,\theta,\phi,t) \, sin\theta \, d\theta \, d\phi ~, \\
    \langle a \rangle (t) = \frac{1}{V} \int a(r,\theta,\phi,t) \, dV~, \quad \Bar{a} = \frac{1}{\Delta t} \int_{t'}^{t'+\Delta t} a(t) \, dt ~.
\end{gathered}
\end{equation*}
We performed the time averages after a stationary state had been reached, and we typically monitored the dimensionless hydrodynamic and magnetic Reynolds numbers, the Rossby number, and the Elsasser number,
\begin{equation*}
\begin{gathered}
    Re = \overline{\sqrt{\langle u^2 \rangle}} ~, \quad  Rm =  \overline{\frac{1}{V} \int_{r_i}^{r_o} \frac{\sqrt{||\mathbf{u}^2||}}{\tilde{\lambda}}r^2 dr } ~, \\
    Ro = \frac{Rm~E}{Pm} ~, \quad \Lambda = \overline{\biggl< \frac{\mathbf{B}^2}{\tilde{\rho} \tilde{\lambda}} \biggl>} ~.
\end{gathered}
\end{equation*}
The total kinetic and magnetic energy, which are in units of $\rho_0 d^5 E^2 \Omega^2$, were
\begin{equation*}
E_{kin} =\frac{1}{2}  \overline{\langle \tilde{\rho} \mathbf{u}^2 \rangle}\,, \quad \quad E_{mag} = \frac{1}{2} \overline{\frac{1}{E Pm} \langle \mathbf{B}^2 \rangle}\,.  
\end{equation*}
We defined the dipole fraction, $f_{dip}=E_{mag,l=1}/E_{mag}$, as the ratio of the magnetic energy stored in dipolar components (axisymmetric and nonaxisymmetric), divided by the total magnetic energy.\footnote{Our definition differs from another widely used definition, that is, the ratio of the axisymmetric dipole component to the magnetic energy in the spherical harmonic degrees $l \, \leq$ 12 at $r_o$, and the total, for instance, \cite{Christensen&Aubert2006}.} 

To study the energy dissipation, we used the buoyancy power,
\begin{equation}
    P_{\nu}(t) \equiv \frac{Ra E}{Pr} \langle  \tilde{\alpha} \tilde{T} \tilde{g} s' u_r \rangle~.
\end{equation}
We used the subindex $\nu$ to emphasize that the quantity was calculated in viscous timescales. For comparison with other works, we also used the rotation timescale (see Sect. \ref{Sec: Scaling laws}). For a well-resolved numerical run, the buoyancy power must be equal to the sum of viscous and Ohmic dissipation rates after a steady-state solution is reached. These are defined as
\begin{equation}
    D_{visc}(t) \equiv \langle S^2 \rangle\,,  \quad  \quad D_{ohm}(t) \equiv \frac{1}{E Pm^2} \langle \tilde{\lambda} \left(\boldsymbol{\nabla}
	\times\mathbf{B}\right)^2 \rangle\,.
\end{equation}
Another quantity to monitor is the fraction of energy dissipated by Joule heating alone, that is, the Ohmic fraction $f_{ohm} = D_{ohm} / P_\nu$. After a statistically steady state has been reached, the input buoyant power must balance the viscous and Ohmic diffusion. To evaluate whether the numerical solution has good time and spatial invariance and also if the background state affects the energy balance strongly, we assessed the power imbalance by its proxy $f_P = |\overline{P_{\nu}} - \overline{D_{visc}} - \overline{D_{ohm}}|/\overline{P_{\nu}}$.

Finally, we studied the time-averaged kinetic and magnetic spectra, that is, the distribution of the energy over different multipoles of order $l$, which MagIC already implemented as a user-friendly output. We inspected the spectra for each model, in particular, to ensure that the resolution was high enough to resolve the maximum dissipation, that is, $l(l+1)E(l)$, for the volume-integrated spectra and 2D spectra taken at relevant radii, for instance, $r_i$, $r_o$, $r_m$. 

\subsubsection{Parameter evolution and model descriptions}
\label{Sec: Parameter evolution}

As explained in \ref{Sec: Background state implementation}, when the MESA profiles were cut close to the desired $N_\rho$, we extracted $\Delta T$, $r_o$, $r_i$, and $r_m$, from which we deduced $\eta$ and $\chi_m :=r_m/r_o$. The corresponding physical values can be recovered by knowing the units in which each quantity is expressed and the values from the MESA profile, for example, using the thickness of the physical shell thickness $d_{phys} = r_{o,phys} (1 - \eta)$. The real quantities of the planet, which reflect the evolutionary changes, were used to evolve the dynamo parameters. Since, as mentioned above, the real physical values $E$ and $Ra$ are computationally inaccessible, we can still use their dependence on the physical values that change during the long-term evolution. In particular, the shell thickness $d=r_o-r_i$ and the temperature difference $\Delta T$ enter the definition of the Ekman, $E(t) \approx d(t)^{-2}$, and Rayleigh numbers, $Ra(t) \approx d(t)^3 \Delta T(t)$. Therefore, we considered a series of ages, for which, after having found a suitable pair of $E_0$ and $Ra_0$ that produces convection and dynamo for the setup with $d_0$ and $\Delta T_0$ corresponding to a given age, the values $E'$ and $Ra'$ of the remaining models in that series were set up by scaling with $d(t)$ and $\Delta T(t)$,
\begin{equation*}
    E' = E_0 \frac{d_0^2}{d'^2}~, \quad Ra' = Ra_0 \frac{d'^3 \Delta T'}{d_0^3 \Delta T_0}~.
\end{equation*}
We made two assumptions: (i) The diffusivities at a given radius remain constant in time, which implies that we considered the same values of $Pr$ and $Pm$ along a sequence and that the change in $E$ only comes from the contraction of the planet; and (ii) planetary rotation is constant in time. The latter assumption is well justified when we consider the possible relevant torque acting on a gas giant. \cite{Batygin2018} studied the evolution of rotation, considering the magnetic coupling between the planetary interior and the quasi-Keplerian motion of the disk in the planetary formation stages. This resulted in efficient braking of the planetary spin that ceased to evolve after $\approx$1 Myr, when it reached a terminal rotation rate (probably similar that of Jupiter), which can hardly change later. Our earliest model is at 100 Myr, for which the rotation can be safely considered constant. In this sense, cold giants are expected to be fast rotators ($Ro<0.12$), and they might host planetary dynamos similar to those found in dipole-dominated numerical solutions with very low $E$ \citep{Davidson2013, Yadavetal2016, Schwaigeretal2019}. In this scenario, there is a quasi-geostrophic balance (Coriolis and pressure forces) at the largest scales, followed by an ageostrophic magneto-Archimedean-Coriolis balance (Coriolis, buoyancy, and Lorentz forces).

With these assumptions, we considered five sets of dynamo models: $(i)$ A long series with a total of 12 evolutionary stages, ranging from 0.1 to 10 Gyr for a 1 $M_J$ planet; $(ii)$ different density ratios, that is, different cutoff radii, for the same 1 $M_J$ model at 1 Gy; $(iii)$ different planetary masses with $N_\rho \approx 3.0$; $(iv)$ several models with $Pm$ and $Pr$ different from 1; and $(v)$ a 4 $M_J$ mass series with $N_\rho \approx 4.6$. When a different mass was chosen, the radial profiles changed (see Fig.~\ref{Fig: CJ profiles conductivities}), so that using eq.~(\ref{Eq: Analytical conductivity Gom10}) with the above-mentioned values of the free parameters $a$ and $\tilde \sigma_m$, we had to adapt the density contrast $N_\rho$ to include the drop in the conductivity in the outer layers of our shell, without at the same time, having too low values of $\sigma$. For this reason, the series of $4\, M_J$ has a higher $N_\rho$ (the exponential drop of $\sigma$ would have been cut out with $N_\rho \approx$ 3). For the same reason, the 0.3 $M_J$ model has a lower contrast, $N_\rho \approx$ 1.1. The alternative would have been to consider an equally arbitrary change of the free parameters in eq.~(\ref{Eq: Analytical conductivity Gom10}). In Table \ref{Tab: Cold Jupiter input parameters} we show the input values for the 3D simulations together with the parameters of the background profiles from the MESA 1D long-term evolution. 

\begin{figure*}[t]
\centerline{\includegraphics[width=\hsize]{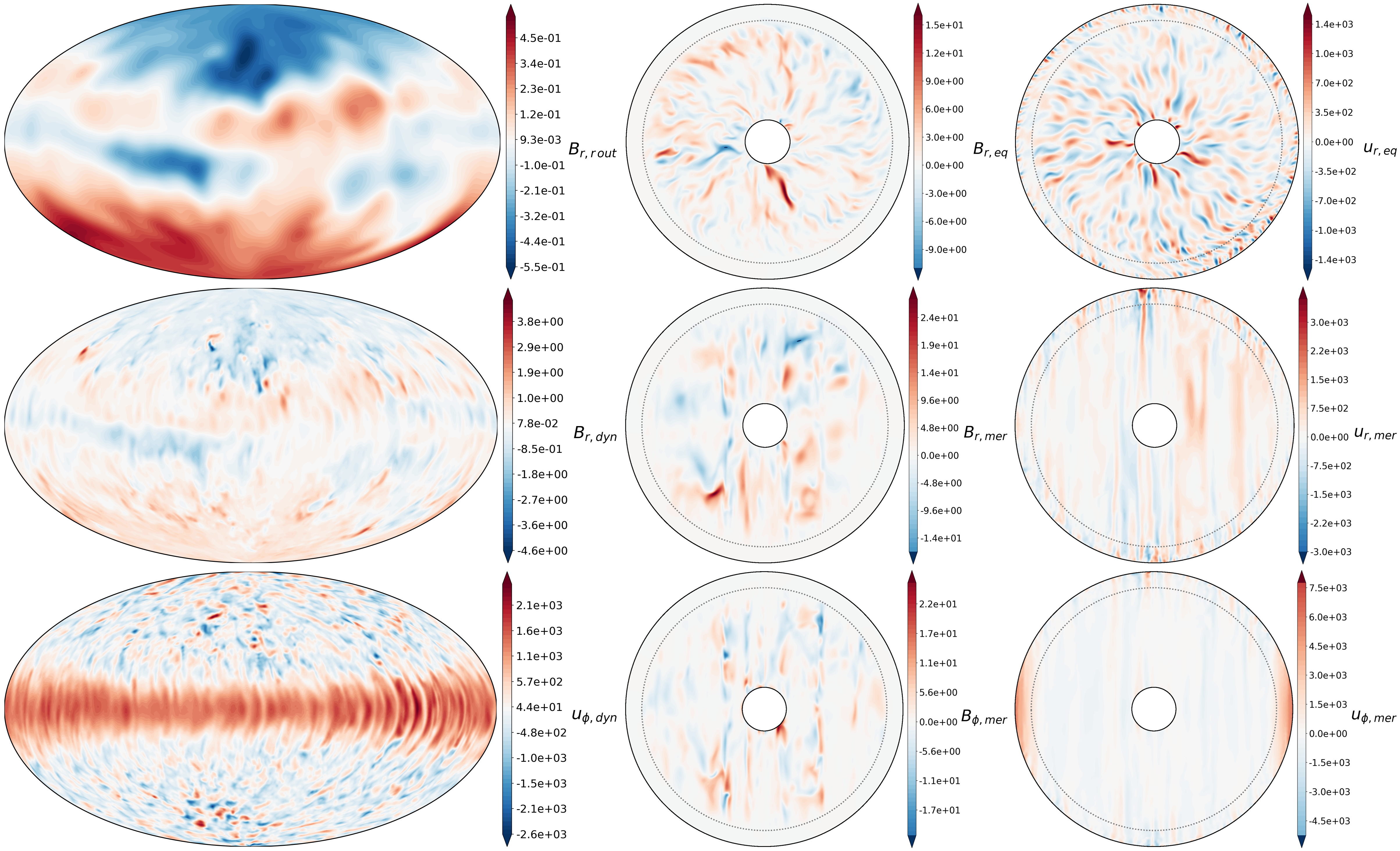}}
\caption{Snapshots of the saturated solution for the representative 1 M$_J$ 1 Gyr model. {\em Left column:} Maps of $B_r$ at the outermost layer of our domain, $B_r$ at $r=r_m$, $u_\varphi$ at $r=r_m$, from top to bottom. {\em Center:} Equatorial slice of $B_r$, meridional slice of $B_r$, and meridional slice of $B_\varphi$. {\em Right:} Equatorial slice of $u_r$, meridional slice of $u_r$, and meridional slice of $u_\varphi$. In the central and right panels, the location of $r_m$ is marked with dotted gray lines in the equatorial and meridional cuts. The color bars indicate the values in code units.}
\label{Fig: 1MJ 1Gy slices}
\end{figure*}

\section{Results}
\label{Sec: Results}

\subsection{Preliminary exploration of parameters}
\label{Sec: Preliminary exploration}

Our first goal was to determine the dependence of the dynamo solutions on mass and age for a given evolutionary sequence of background thermodynamic setups (Table \ref{Tab: Cold Jupiter input parameters}). Therefore, the first requirement was to determine a range of parameters for which both convection and dynamo operated for all models. This practically meant that we needed to find a feasible range of $Ra$ and $E$ (which, within a sequence of models, have relative variations set by $\Delta T$, $T_o$ and $d$, Table \ref{Tab: Cold Jupiter input parameters}), for which convection and dynamo action are present in the entire sequence, keeping in mind that the chosen values were orders of magnitude away from the realistic values, as mentioned above. To determine these feasible ranges, we needed to perform a preliminary parameter exploration. At the same time, we assessed the sensitivity of results on other parameters. We summarize in this subsection the four main steps we took for this overall assessment.

First, to locate the region with viable dynamo solutions, we performed low- and medium-resolution runs for one specific model, the model with 1 $M_J$ 10 Gyr. This is the oldest and coldest model of the 1 $M_J$ sequence, that is, with the lowest value of $\Delta T$, which also implies the lowest value $Ra$ of the series, that is, it is least favorable to convection. We spanned the ranges 10$^{-5}$ $<E<$ 10$^{-3}$, 10$^{6}$ $<Ra<$ 10$^{10}$ with $Pm$=$Pr$=1 and a relatively low resolution, ($N_r$,$N_{\theta}$,$N_{\phi}$) = (193,192,384). For $E \le$ 10$^{-4}$, we obtained convection for $Ra \gtrsim 10^{7}$, and, additionally, magnetic field growth for $Ra \gtrsim 5\cdot$10$^{7}$.

Second, for the $1~M_J~10$~Gyr case with $E$=10$^{-5}$, $Ra$=5$\cdot$10$^{8}$ (the reference model in the rest of this subsection), we explored the Prandtl numbers in a relatively easily accessible range 0.25 $<Pm, Pr<$ 4. This was done to discuss the impact of our assumptions of constant-in-time diffusivities Sect. \ref{Sec: Transport coefficients}).  The results of this exploration are shown in Sect. \ref{Sec: Dependence on Prandt number} for high-resolution models and different evolutionary ages.

Third, for the same reference model ($E$=10$^{-5}$, $Ra$=5$\cdot$10$^{8}$, $Pm$=$Pr$=1), we explored the sensitivity to the parameters $a$ and $\sigma_m$, which define the slope of $\tilde{\lambda}(r)$ in the outermost layers. For a wide range of values (i.e., 0.07 $<\sigma_m<$ 0.9, 5 $<a<$ 15), we indeed recovered the results of \cite{Duarteetal2013}, that is, the strong equatorial jet remains confined to the weaker conducting outer region and does not interfere with the deeper dynamo action. As previously mentioned, we finally opted for rather low values for both $\sigma_m$ of 0.1 and $a$ of 7. These values are similar to those used by \cite{Gastine&Wicht2021}, whose model approximately reproduces the \cite{Frenchetal2012} profiles. For numerical stability reasons, we did not use higher values of $a$, that is, steeper exponential drops, because the hydrogen metallic region of some of our models is already quite deep ($\chi_m>$0.85), which leads to too high values of $\tilde{\lambda}(r)$ near the outer surface. Similarly, we studied the effect of $N_\rho$ on several diagnostic quantities (see App.~\ref{App: Sensitivity on the density ratio} for more details). Given the non-negligible effects of choosing different values of $N_\rho$, we compare below models with the same $N_\rho$, to avoid additional biases.

Fourth, we tested different boundary conditions again for the same reference model. Integrated quantities such as $Rm$ or $E_{kin}$ and convection patterns did not change appreciably when rigid boundary conditions were applied at the inner core. The radial distributions showed a drop in velocity in a very thin region ($<1\%$) of the radius. For the magnetic field, we tested for a perfect conductor and insulator for the inner core and an insulating, perfect conductor and pseudo-vacuum at the outer radii. We found no relevant differences in the internal dynamo.

Considering the explored values of $Ra$ and $E$ for these low-resolution test, we set $Ra=$ 1.3$\cdot10^{-5}$ and $E \approx 5\cdot10^{8}$ for the 1 $M_J$ 10 Gyr model with a $N_\rho \approx$ 3. This choice set the rescaled values for the remaining sequence in a range that allowed dynamo action. Using the definition of $E$, $Ra$, we scaled their values in each run with the corresponding values of $T$, $\Delta T$, $d$, as described in Sect. \ref{Sec: Parameter evolution}. Most models use a resolution of ($N_r$,$N_{\theta}$,$N_{\phi}$) = (289,256,512). As an exception, the 4 $M_J$ $N_\rho \approx 4.6$ runs use a grid of (385,320,640).

Finally, we employed a strategy to save computational time that was inspired by other spherical shell dynamo works \citep{Christensen&Aubert2006}. The idea is that the final solution does not depend on whether the initial conditions are taken from another saturated dynamo model or if they are the usual $\mathbf{u}=0$ with a weak perturbation of both $\mathbf{B}$ and $T'$/$s'$. This suits our case in particular because the relative changes in the dynamo parameters from one setup to the next are small, and the solution of the new setup is reached much faster than starting from a $\mathbf{u}=0$ state. In App. \ref{App: Continuing models from past models} we show the results of some specific numerical experiments that support this strategy.

We now proceed to the main results. We refer to App.~\ref{App: Tabulated runs} for a table with detailed values of the time-averaged output nondimensional numbers and the other quantities we used as diagnostics.

\begin{figure*}[t]
\centering
\includegraphics[width=.495\textwidth]{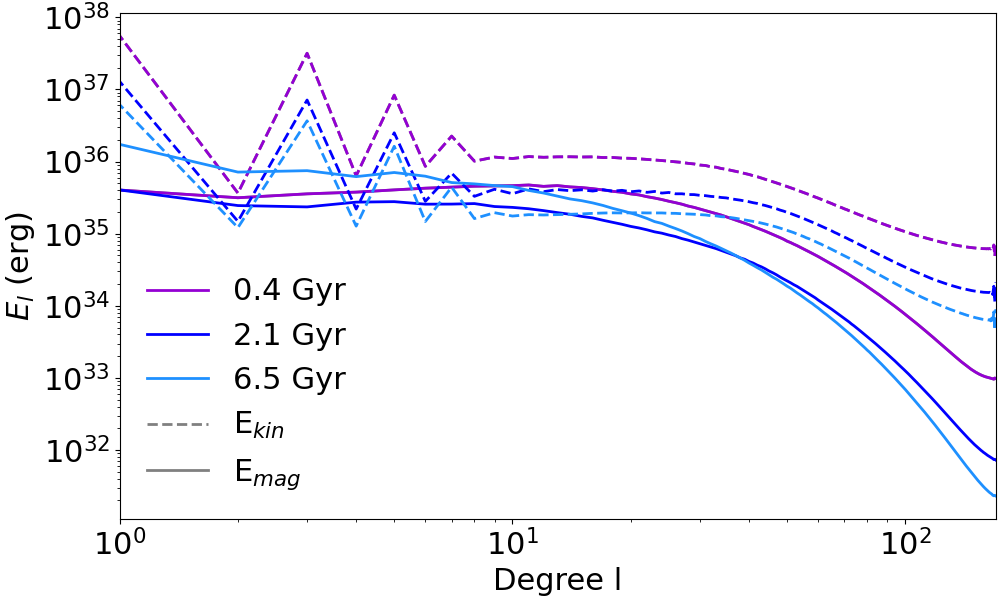}
\includegraphics[width=.495\textwidth]{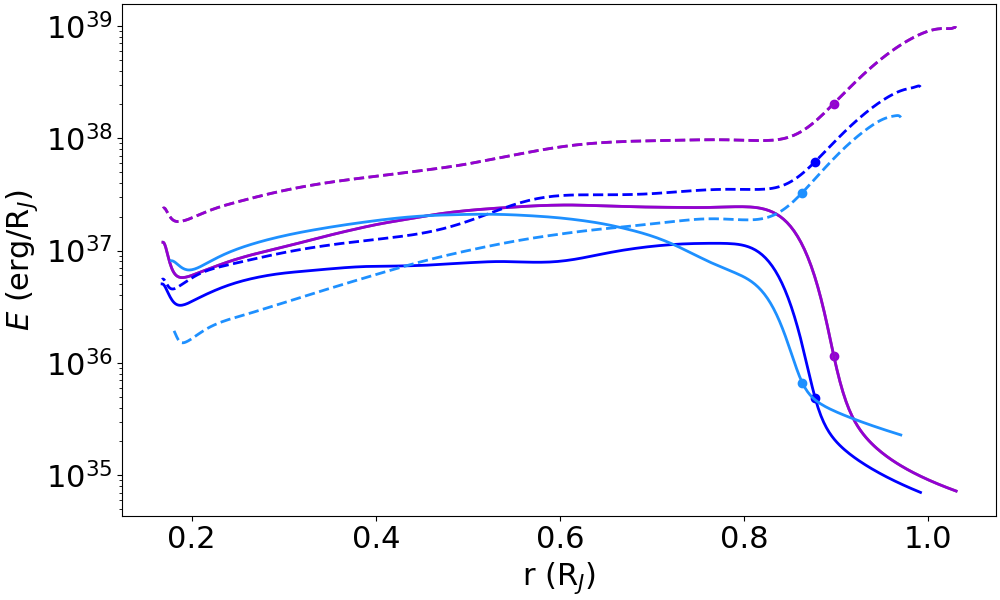}
\caption{Magnetic (solid) and kinetic (dashes) energy distribution over the multipole degrees $l$ (left), and over the radius (right) for three models representing the same 1 $M_J$ planet at different evolutionary stages (0.4, 2.1, and 6.5 Gyr). The spectra have been averaged in time over the saturated state. The location of $r_m$ is marked with a dot in the radial plots. The physical units are obtained by multiplying by the factor $\rho_0 d^5 E^2 \Omega^2$, where $\rho_0$, $d$ and $E$ depend on each model and $\Omega = 1.76\cdot 10^{-4}$, i.e., the Jovian value.}
\label{Fig: Energy spectra and radial distribution for 1MJ OK models}
\end{figure*} 

\subsection{Dynamo solutions: General behavior}

In Fig.~\ref{Fig: 1MJ 1Gy slices} we show snapshots (maps and slices of some velocity and magnetic field components) of the 1 M$_J$ 1 Gyr saturated dynamo solution, which is a representative case. The other models we obtained are qualitatively similar among themselves in terms of the morphology of the velocity and magnetic fields. The largest differences are the relative average strengths of $\mathbf{u}$ and $\mathbf{B}$ and the magnetic field dipolarity, which we discuss below. All the models have a strong equatorial flow that reaches deep, down to the dynamo region. The velocity and magnetic fields show a westward drift in the inner parts. The magnetic field is mostly constrained under $r<r_m$, where convection is also stronger, as seen in $u_r$. As expected from rotation-dominated convection, we found columnar structures in the direction of the rotation axis \citep{Zhang&Busse1987, Ardesetal1997, Simitev&Busse2003}, as shown in the meridional slices. Generally speaking, our numerical solutions are similar to those reported by other works for Jovian-like dynamos with a nonconstant electrical conductivity \citep{Jones2014, Duarteetal2018}.\footnote{They use a different code with a strictly isentropic background profile that fits $T(r)$, $\rho(r)$ and $\sigma(r)$ from \cite{Frenchetal2012}.}

The dynamo solutions we found are generally less dipole-dominated than their incompressible (Boussinesq) counterparts with similar dynamo parameters (i.e., $Ra$, $E$, $Pr$, and $Pm$). However, we note that we did not explore values for $Pr$ lower than 0.1 with $Pm>1$, where dipole-dominated solutions have been found \citep{Jones2014, Tsang&Jones2020}. We restricted ourselves to a less demanding parameter space with a wider liberty of parameter exploration, but with the caveat that we might obtain less dipole-dominated models. Similarly to \cite{Yadavetal2013}, for all runs shown here, we also obtained $Nu>2$ at the top and bottom surfaces, which ensures a fully developed convection. The Nusselt number $Nu$ is the ratio of the total transported heat flux to the conducted heat flux.

Moreover, for gas giants, the definition of the dynamo surface is not absolute. As hydrogen gradually transitions outward from metallic to molecular, the electrical conductivity and electrical currents are quickly (but not abruptly) damped over a finite region. For our models, the most obvious choice for the dynamo surface is the radius at which the exponential decay for $\sigma$ starts, that is, $r_m$. To obtain a more physically justified definition for the dynamo surface, we followed \cite{Tsang&Jones2020} by computing the magnetic energy spectra at different radii, $F_l(r)$. They defined the dynamo surface, $r_{dyn}$, as the radius within which the slope of the Lowes spectrum (i.e., a potential solution extrapolated back from the outermost layer to the interior) diverges from the slope of the simulated $F_l(r)$. A similar analysis for some of our models is shown in App.~\ref{App: Dynamo surface definition}, where we find that this definition of the dynamo surface always gives values very close to $r_m$.

\subsection{Evolutionary changes}
\label{Sec: Evolutionary changes}

\begin{figure*}[t]
\centerline{\includegraphics[width=\hsize]{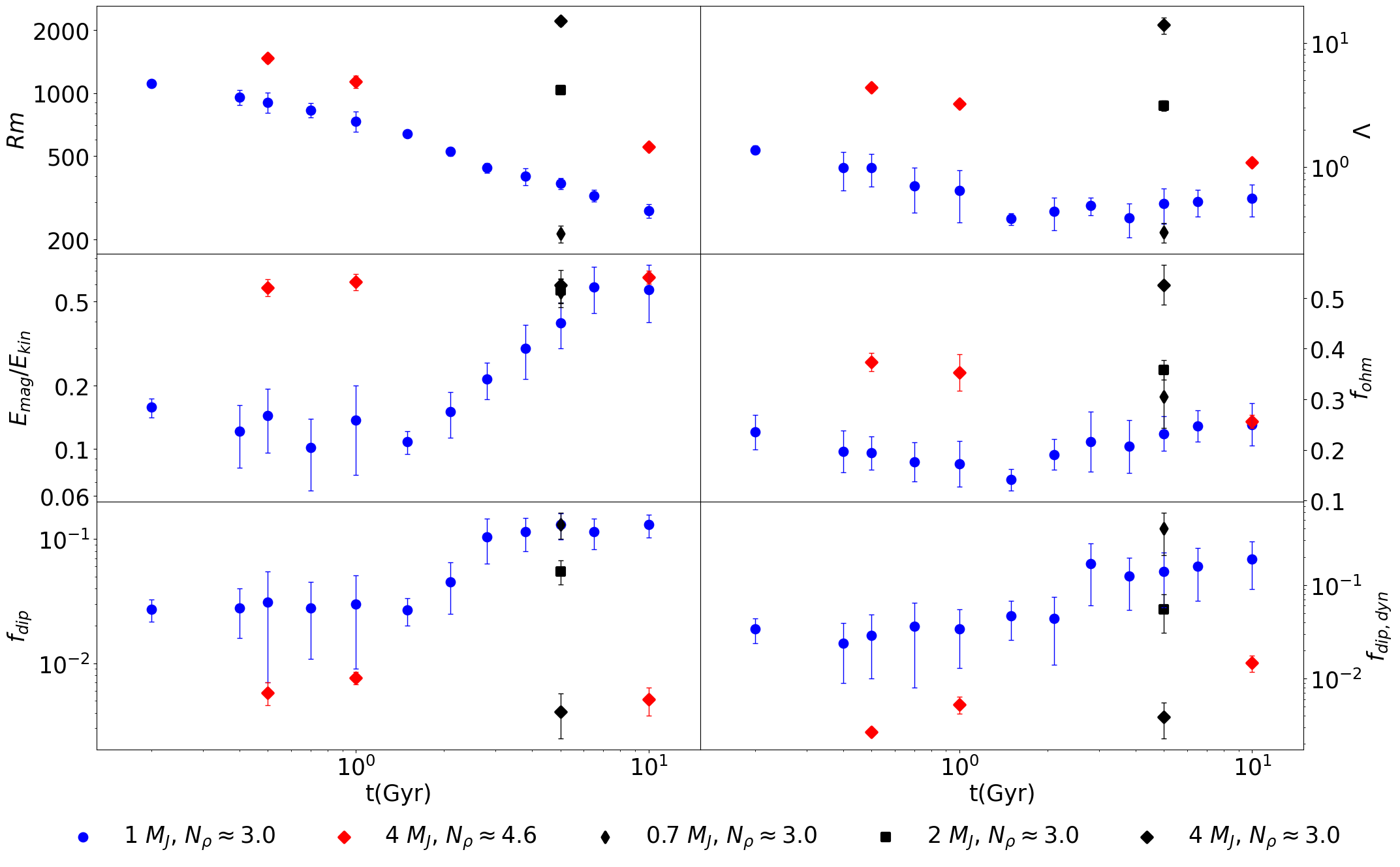}}
\caption{Diagnostics as a function of the age for different masses, all with $Pm=Pr=1$. From left to right and top to bottom, we show the magnetic Reynolds number, the Elsasser number, the magnetic-to-kinetic energy ratio, the Ohmic fraction, total dipolarity, and dipolarity at the dynamo surface. The 4 $M_J$ runs at 0.5, 1, and 10 Gyr have a higher $N_\rho$ (see text).}
\label{Fig: General diagnostics}
\end{figure*}

We now focus on the variation in the solutions along the longest sequence of models, the 1 $M_J$ planet with $N_\rho\approx3$ and $Pm=Pr=1$. The shell-averaged spectral distribution and radial distribution of the magnetic (solid lines) and kinetic (dashed lines) energy are shown in Fig. \ref{Fig: Energy spectra and radial distribution for 1MJ OK models}. We focus on three representative ages: 0.4, 2.1, and 6.5 Gyr. 

In the left panel, the kinetic spectra show a drop of about 1.5 orders in magnitude or more from the integral to viscous scales, while the magnetic spectra decrease by at least 3 orders. The sawtooth shape on the lowest multipole side is associated with the external jet that we see in all of our runs. This behavior is not seen for the $m$ spectrum, which is dominated by the zonal flows, $m=0$. For higher multipoles, the spectrum plateaus before it reaches the viscous scale and drops off. A comparison of the three different models shows that the overall shape of the kinetic spectra does not significantly change, other than a constant decrease in time throughout all harmonic degrees. The magnetic spectra show a similar diffusive scale, which is approximately located at the same $l$ as the viscous diffusive scale (compatible with $Pr=Pm=1$), although the knee is less pronounced. With the usual measure of dipolarity, it ranges from $0.3 > f_{dip,l<12}^{axi, surf} = E_{mag}(r_0)_{l=1,m=0}/E_{mag}(r_0)_{l\leq12} > 0.8$. Depending on the work, this could be considered multipolar or dipolar \citep{Christensen&Aubert2006, Yadavetal2013, Zaireetal2022}. The magnetic spectra show a clear evolution with age: The relative weights of high multipoles tend to decrease, while the strength of the large scales slightly increases. This inversion leads to an increase in the total dipolarity (see the discussion below).

The radial energy distributions, shown in the right panel of Fig.~\ref{Fig: Energy spectra and radial distribution for 1MJ OK models}, shows that $E_{kin}(r)$ increases almost monotonically outward, but the steepest changes occur in the outermost layers, $r \gtrsim r_m$, due to the appearance of the equatorial zonal wind, where the density is lower and the magnetic drag is weaker than in the interior. Similarly to the spectra, the kinetic radial distribution does not show a clear variation with age. The radial profile $E_{mag}(r)$ and its change with age shown also in Fig.~\ref{Fig: Energy spectra and radial distribution for 1MJ OK models} is instead more complex. The radial profile peaks at radii slightly smaller than $r_m$, after which it significantly drops, following the $\sigma(r)$ profiles (Fig.~\ref{Fig: CJ profiles conductivities}). A comparison of different evolutionary ages shows that the innermost region of the radial distribution does not show a clear trend, with a slight increase in the deepest regions for late-age models. On the other hand, the layers $\approx$ 10-20 \% below $r_m$ show a steady decrease with age (see Sect. \ref{Sec: Magnetic field strength at the dynamo surface}).

The overall changes between the runs are gradual, and we found that a few models can already predict the general behavior. Several time and volume-averaged diagnostic quantities are shown as a function of evolutionary models in Fig. \ref{Fig: General diagnostics}. As expected during the planetary cool-down of a gas giant, $Rm$ decays approximately like a power law in time. $Ro$ and $P_\nu$ also behave similarly, and they are therefore not shown to avoid repetition. All of these quantities are dependent on $u_{rms}$ or at least on one of its components. This reflects the fact that the mean velocity is dictated by the buoyancy input parameter $Ra$, which is proportional to the temperature difference in the convective shell. Fig. \ref{Fig: General diagnostics} also shows that the dipolarity $f_{dip}$ does not seem to change significantly in time, except for a mild increase between 1.5 and 3.8 Gyr. This appears to indicate a transition between a multipolar or weakly dipolar-dominated regime to a strong dipolar-dominated regime. This increasing trend is obscure, with $f_{dip,l<12}^{axi, surf}$ (for the 1 $M_J$ series, they take values from 0.3 to 0.8). As an alternative, we obtained the average $f_{dip}$ over the volume 5\% near the dynamo surface, $f_{dip,dyn}$. Fig. \ref{Fig: General diagnostics} shows that $f_{dip,dyn}$ grows gradually, with a similar jump seen in $f_{dip}$.

\begin{figure}[t]
\centerline{\includegraphics[width=\hsize]{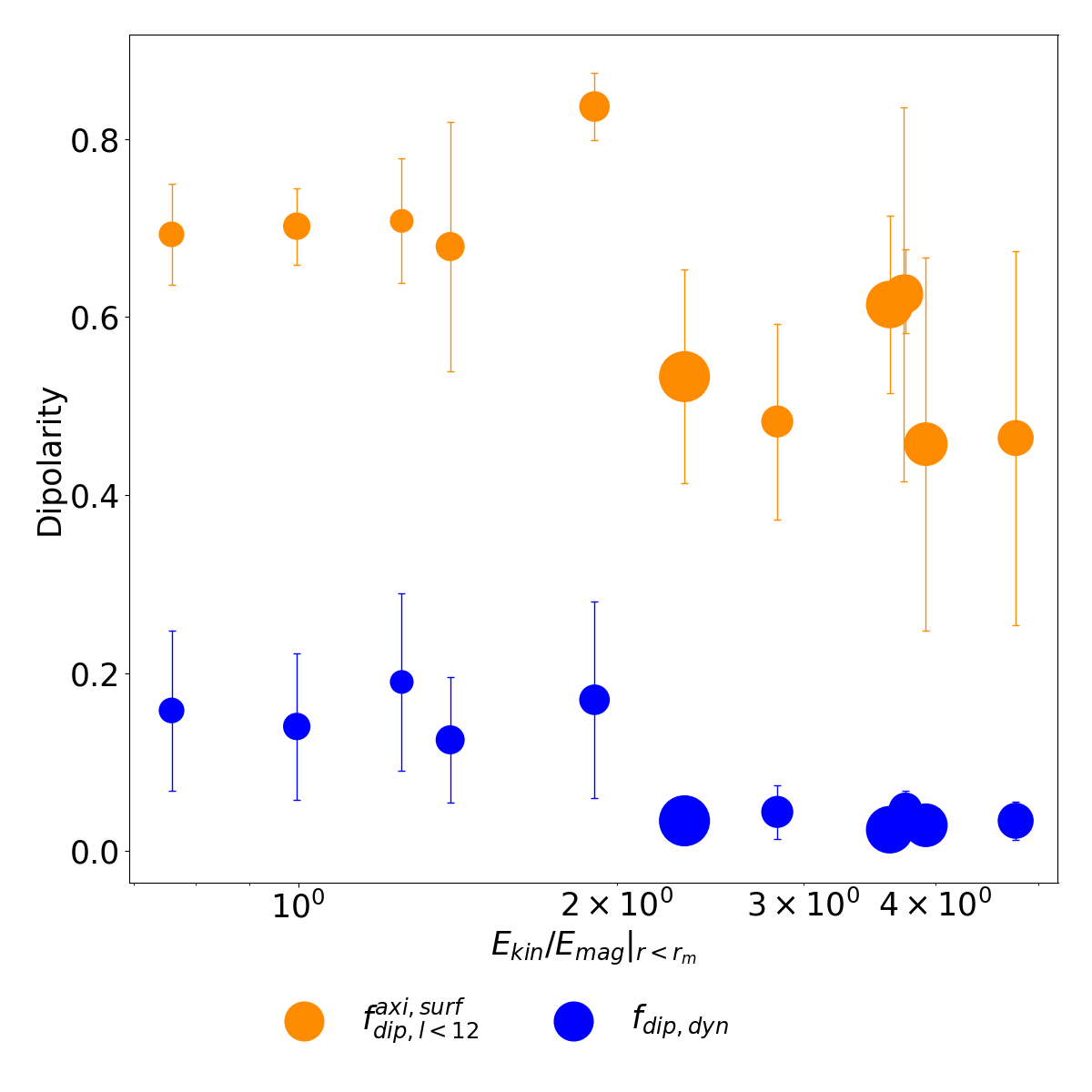}}
\caption{Dipolarity measurements as a function of the inverse of equipartition for the 1 $M_J$, $Pr=Pm=1$ models.}
\label{Fig: fdip vs equip}
\end{figure}

This transition from multipolar to dipolar dynamo was also observed by \cite{Zaireetal2022} for dynamos in stratified stellar interiors, which they modeled with shallower shells ($r_i/r_o=0.6$) and different $N_\rho$ and $Ra$. They obtained a threshold $F_I/F_L$ (i.e., the relative importance of the inertial over Lorentz forces) below which multipolar dynamos collapse into dipolar dynamos. They also reported that $E_{kin}/E_{mag}$ can equally well capture this magnetic morphology transition and found this transition at about $E_{kin}/E_{mag}$=0.7. In Fig. \ref{Fig: fdip vs equip} we show both $f_{dip,l<12}^{axi,surf}$ and $f_{dip,dyn}$ as a function of $E_{kin}/E_{mag}$ for the 1 $M_J$, $Pr=Pm=1$ series. We also report two distinctive populated areas, low dipolarity with high $E_{kin}/E_{mag}$ and high dipolarity with lower $E_{kin}/E_{mag}$. This abrupt change in the magnetic field morphology seems to be better reflected with $f_{dip,dyn}$. A good definition for a dipole-dominated dynamo could be $f_{dip,dyn}>0.1$, which might be compatible with the definitions of \cite{Yadavetal2013} or \cite{Zaireetal2022} of $f_{dip,l<12}^{axi,surf}>0.3$ and $0.5$, respectively. These dipole-related quantities are usually the most fluctuating integrated quantities in saturated dynamo solutions because they are susceptible to the specific magnetic field configuration. 

In any case, our strongest evolutionary trend shows a transition from a multipolar to a dipolar regime in the middle of our series. Within the more multipolar part of the series, $\Lambda$ and $f_{ohm}$ show a minor decay in time, and $E_{mag}/E_{kin}$ also seem to decrease, but more subtly. In the dipolar regime, these trends reverse: $\Lambda$ and $f_{ohm}$ plateau and show a slight increase; the ratio $E_{mag}/E_{kin}$ grows noticeably. Following the magnetic fields of Jupiter and Saturn \citep{Connerney2022, Caoetal2023}, gas-giant dynamos are expected to live in a parameter space region with dipole-dominated solutions, and thus, the expected evolution would be of the latter part of our series. Moreover, the Jovian value for $Pm$ is expected to be $\approx 10^{-6}$, meaning that $D_{ohm} >> D_{visc}$, and thus, $f_{ohm}$ is expected to be close to one. This would mean that our predicted $f_{ohm}$ trend is not physically noticeable. An increase in $E_{mag}/E_{kin}$ and $\Lambda$, the integrated nondimensional magnetic energy, might contradict the aforementioned scaling laws, but we show their compatibility in Sect. \ref{Sec: Magnetic field strength at the dynamo surface}.

\begin{figure*}[t]
\centering
\includegraphics[width=.495\textwidth]{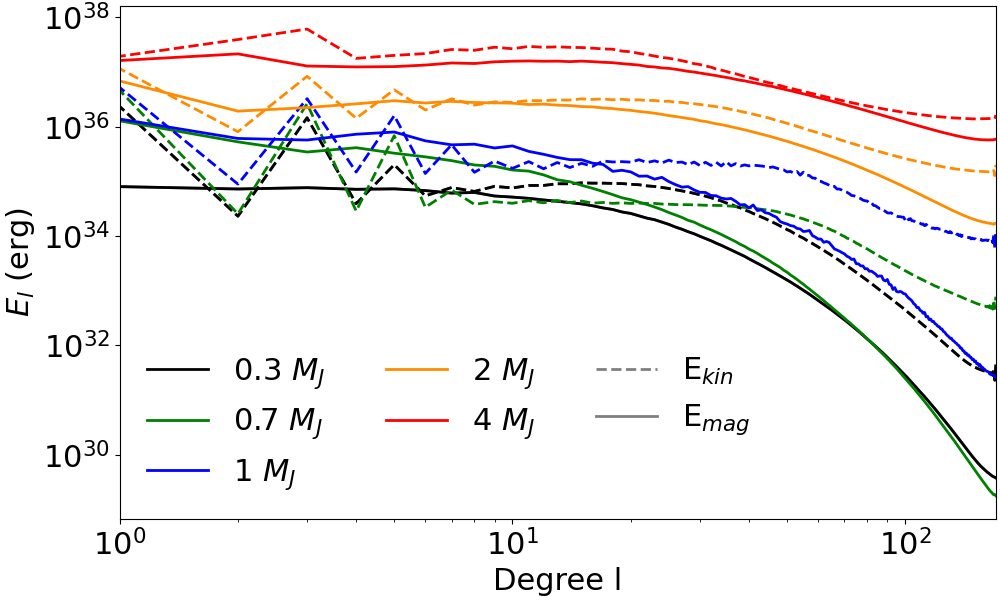}
\includegraphics[width=.495\textwidth]{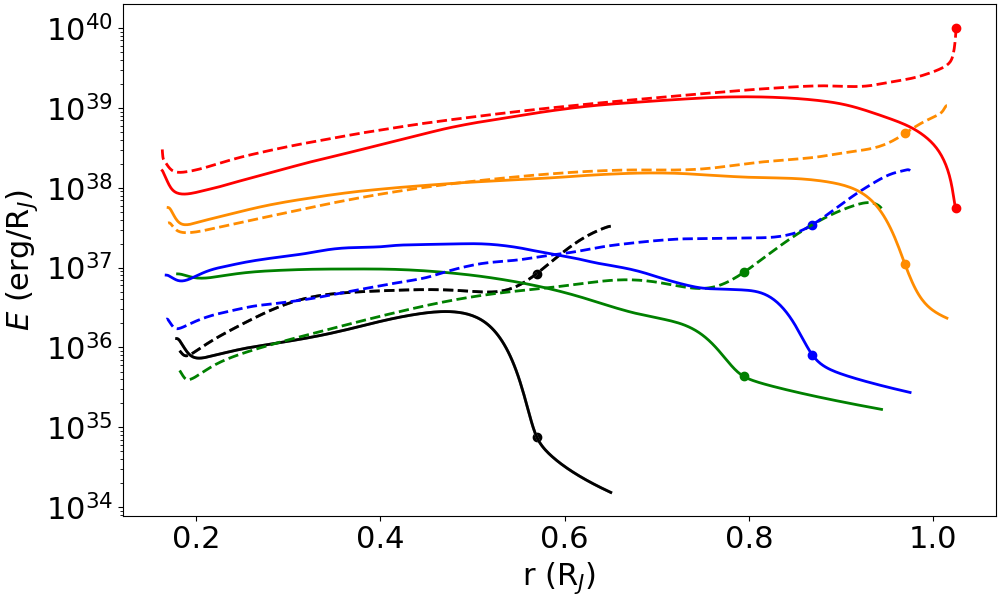}
\caption{Magnetic (solid) and kinetic (dashed) energy distribution over the multipole degrees $l$ (left) and over the radius (right) for planets with different masses at 5 Gyr. The location of $r_m$ is marked with a dot in the radial plots.}
\label{Fig: Energy spectra and radial distribution for different mass models}
\end{figure*}

The values of $E_{mag}/E_{kin}$ that we show are below equipartition (i.e., $0.1<E_{mag}/E_{kin}<0.6$ for the long 1$M_J$ sequence) because we analyzed a volume including the nonconducting outer layer. When we restrict the energy integration within the metallic region $r<r_m$, then $0.25<E_{mag}/E_{kin}|_{r<r_m}<1.4$. This tendency can be sensed from the radial distributions in Fig. \ref{Fig: Energy spectra and radial distribution for 1MJ OK models} and \ref{Fig: Energy spectra and radial distribution for different mass models}.

\subsection{Dependence on planetary mass}

We have obtained saturated models for five different masses at 5 Gyr. We directly compared only four of them because the conductivity of the 0.3 $M_J$ model was not numerically feasible with $N_\rho\approx3$ (see above). The model of 4 $M_J$ is slightly under-resolved, possibly because of its higher $Ra$ and a very little conductivity drop, which does not help to stabilize the stress-free boundary conditions. We did not use higher-mass 1D models (specifically, the 8 and 12 $M_J$) because of the resolution constraints required by the $E$, $Ra$ combination.

The main dimensionless diagnostics for these runs were already shown in Fig.~\ref{Fig: General diagnostics}. The overall trends are dictated by the increase of $Ra$ with a decrease in $E$ that comes with the 1D profiles themselves. Therefore, as the mass increases, both $f_{dip}$ and $f_{dip,dyn}$ decrease, while $f_{ohm}$, $\Lambda$, and $Rm$ increase. The energy ratio is approximately maintained.

In Fig.~\ref{Fig: Energy spectra and radial distribution for different mass models} we plot the energy spectra and radial distribution for these models. The key features are very similar to the feature described above (Fig.~\ref{Fig: Energy spectra and radial distribution for 1MJ OK models}). A noticeable difference is that the magnetic spectra seem to become flatter for higher masses. The sawtooth shape in the kinetic spectra diminishes with mass because depth of the outer nonconductive layer decreases as the hydrogen metallization pressures are reached faster. To overcome this difference and to observe whether the evolutionary trends shown in Fig. \ref{Fig: General diagnostics}  changed for other masses, we obtained saturated dynamos for the 4 $M_J$ 0.5, 1, and 10 Gyr $N_\rho\approx4.6$ models. For these three, we obtained similar dynamos with larger equatorial jets, in other words, we recovered the sawtooth shape seen in the kinetic spectrum. The evolutionary trends match the multipolar side of the 1 $M_J$ long series.

\subsection{Evolution of the magnetic field strength at the dynamo surface}
\label{Sec: Magnetic field strength at the dynamo surface}

\begin{figure}[t]
\centerline{\includegraphics[width=\hsize]{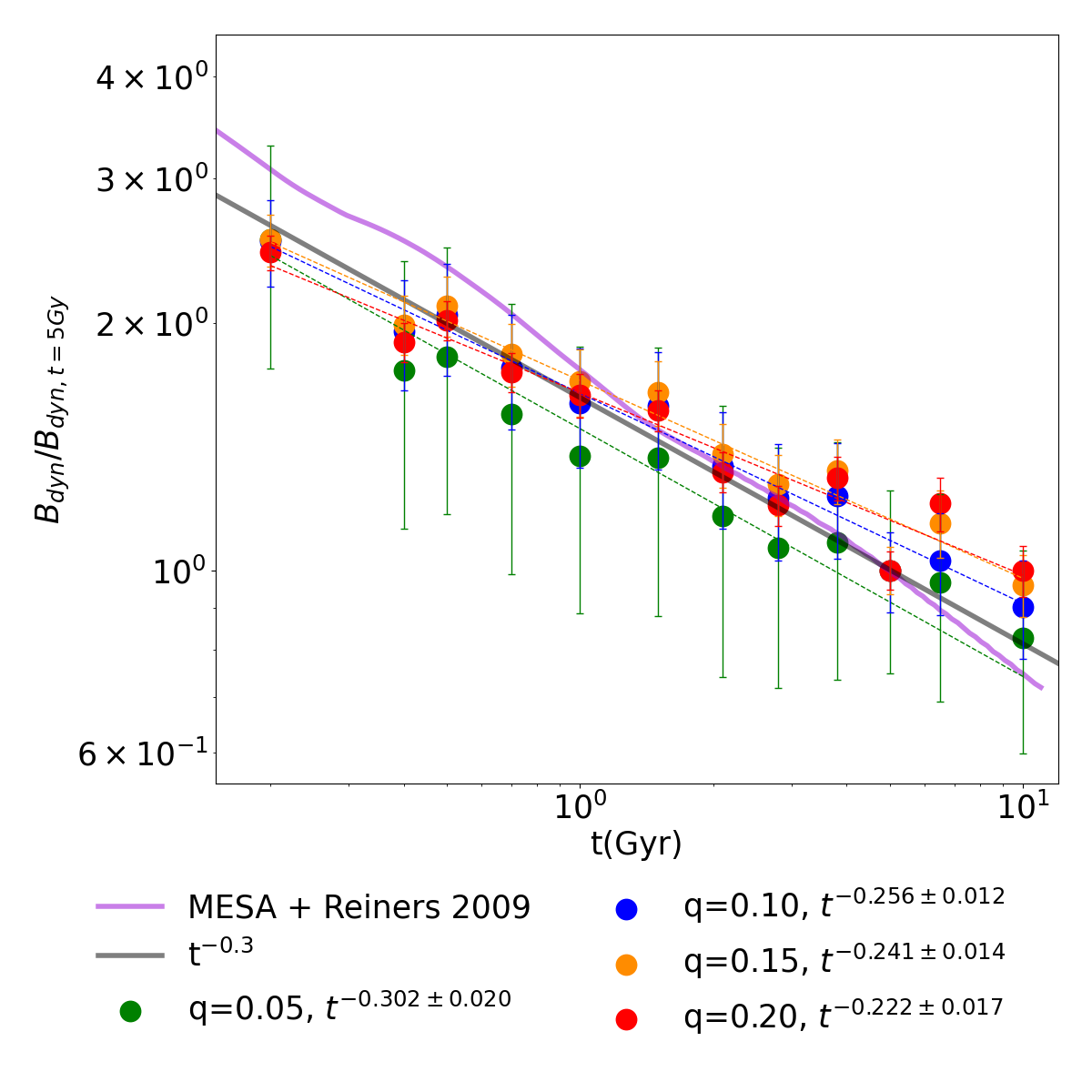}}
\caption{Evolution of the magnetic field strength at the dynamo surface averaged in time, using the scaling laws and calculating the average value of the magnetic field over different relative thicknesses $q$, around $r_m$ (green, blue, orange,  and red). The error bars are associated with the radial variation and are larger for thinner integration shells over which we evaluated eq.~\ref{Eq: Emag dyn}. The dotted lines show the corresponding best-fit power laws. The solid lines indicate eq.~(\ref{Eq: Bdyn scaling law}) applied to our MESA output (pink), and the prediction by \cite{Reiners&Christensen2010} (gray).}
\label{Fig: Bdyn vs time}
\end{figure}

The scaling law provided by \cite{Reinersetal2009} gives the mean strength of the magnetic field at the dynamo surface, $B_{dyn}$, in terms of the mass $M$, luminosity $L$, and radius $R$ of the substellar object,
\begin{equation}
    B_{dyn} = 4.8_{-2.8}^{+3.2} \left( \frac{M}{M_{\odot}} \right)^{1/6} \left( \frac{L}{L_{\odot}} \right)^{1/3} \left( \frac{R_{\odot}}{R} \right)^{7/6}~{\rm kG}~.
    \label{Eq: Bdyn scaling law}
\end{equation}
We evaluated $B_{dyn}$ by using the $L(t)$, $R(t)$ output from our MESA simulations (pink lines in Fig.~\ref{Fig: Bdyn vs time}).

Another slightly different estimate comes from inserting the analytical expressions for $L(t)$ and $R(t)$ of \cite{Burrows&Liebert1993,Burrowsetal2001}  in eq.~(\ref{Eq: Bdyn scaling law}) as given for a substellar-mass solar-metallicity object,
\begin{eqnarray}
    L &\approx& 4\cdot 10^5 L_{\odot} \left( \frac{1 Gy}{t} \right)^{1.3} \left( \frac{M}{0.05 M_{\odot}} \right)^{2.64}~,\\
    R &\approx& 6.7\cdot 10^4 km \left( \frac{10^5 \, {\rm cm} \, {\rm s}^{-2}}{g} \right)^{0.18} \left( \frac{T_{eff}}{1000 K} \right)^{0.11}~. \label{Eq: B Burrows}
\end{eqnarray}
Using these estimates, \cite{Reiners&Christensen2010} obtained a slow decay of the dynamo magnetic field of about one order of magnitude in about 10 Gyr. Fig. 1 in their paper shows that the approximate power-law relation is $B_{dyn} \approx t^{-0.3}$ (marked with a gray line in Fig.~\ref{Fig: Bdyn vs time}).

We then compared these methods with ours. To do this, we evaluated the values of $B_{dyn}$ for our models. As mentioned above, we made use of the fact that the effective dynamo surface is located at $r_m$. We decided to obtain the volume average of $E_{mag}$ over a spherical shell from $r_m$ to some not-too-deep layer,
\begin{equation}
\begin{aligned}
    E_{mag,dyn}(q) = \frac{1}{r_m-r_m'} \int_{r_m'}^{r_m} E_{mag}(r) dr~,\label{Eq: Emag dyn}
\end{aligned}
\end{equation} 
where $r_m'(q)=(\chi_m-q)r_o$, and $E_{mag}(r)$ was obtained from the previously shown radial distributions. The shell thickness is therefore controlled by the parameter $q$, which in turn allowed us to evaluate the surface dynamo field as $ B_{dyn}(q) = \sqrt{2 \mu_0 E_{mag,dyn}(q)/M V }$. In Fig. \ref{Fig: Bdyn vs time} we show the estimated $B_{dyn}$ for $q$ = 0.05, 0.1, 0.15, and 0.2 (colored points in Fig.~\ref{Fig: Bdyn vs time}, with the related statistical error). The best-fitting slope (dotted lines) decreases with thicker integrating regions, that is, larger $q$. Within the standard deviations, we recover the previously mentioned slope of $\approx t^{-0.3}$ for the most uncertain slope (thinnest averaging region). The others are slightly shallower than the trends obtained by \cite{Reiners&Christensen2010} and \cite{Reinersetal2009}.

\begin{figure}[t]
\centerline{\includegraphics[width=\hsize]{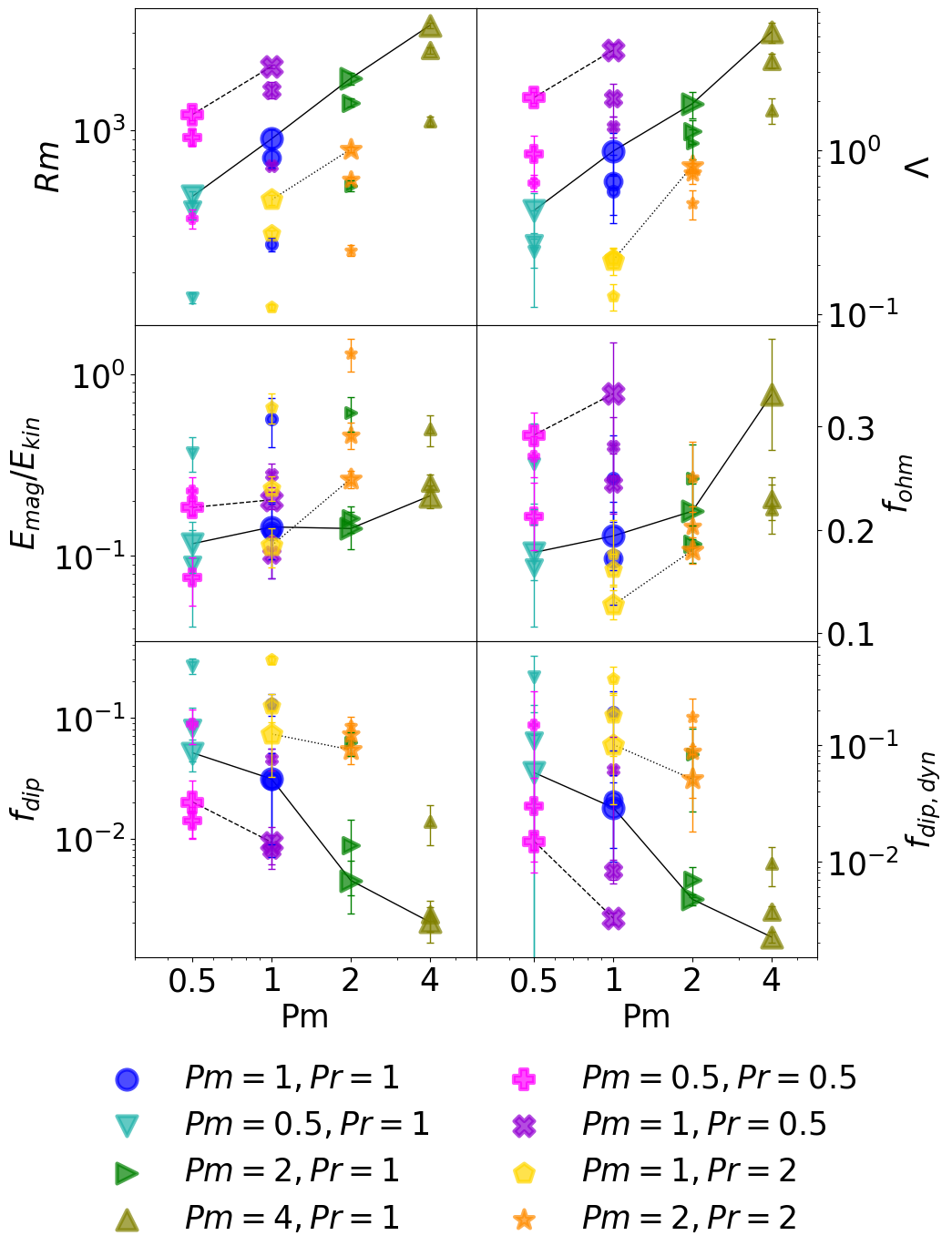}}
\caption{Same diagnostics as in Fig. \ref{Fig: General diagnostics}, shown as a function of $Pm$, with different values of $Pr$. The decreasing size of the mark indicates the increase in age (0.5, 1, and 10 Gyr). The colors and shapes help to distinguish the evolutionary changes and are the same as in Fig. \ref{Fig: Ro_vs_PPm} and \ref{Fig: Losqrtfohm_vs_PPm_&_taumag_vs_Ro}. The lines of constant $Pr$ were added for the 0.5 Gyr models.}
\label{Fig: General diagnostics Pm}
\end{figure}

\subsection{Dependence on the Prandtl numbers}
\label{Sec: Dependence on Prandt number}

To assess the impact of the assumption of constant $Pr$ and $Pm$, we obtained several saturated dynamo states with $Pr, Pm\neq1$ for some 1 $M_J$ models. We performed runs with 0.5, 1, and 10 Gyr, which we deemed enough for assessing general properties of the trends. We investigated within the following range: 0.5 $<Pm<$ 4 and 0.5 $<Pr<$ 2. The results are shown in Fig. \ref{Fig: General diagnostics Pm}.

The increase in $Pm$ can be understood as lowering $\lambda$ while keeping $\nu$ constant, or similarly, increasing $\nu$ with constant $\lambda$. Both effects lead to an increase in the magnetic energy in relation to the available buoyant power. Therefore, Fig. \ref{Fig: General diagnostics Pm} shows that $Rm$ and $\Lambda$ tend to increase with $Pm$, which is a more efficient dynamo mechanism \citep{EliasLopezetal2024}. The same applies for $E_{mag}/E_{kin}$ and $f_{ohm}$, as the decreasing $\lambda$ increases the magnetic energy percentage and the Ohmic dissipation contribution. The surface and volumetric dipolarities both decrease with increasing $Pm$, which is compatible with what was found by \cite{Tsang&Jones2020}: A higher $Pm$ means a less steep magnetic spectrum, or in other words, a more weakly dipole-dominated magnetic spectrum.

In contrast, $Rm$, $\Lambda$, and $f_{ohm}$ decrease for increasing $Pr$. This can easily be understood by considering a higher $Pr$ as increasing values of $\nu$ in comparison to $\kappa$. A higher viscosity will lead to lower kinetic as well as magnetic energy, and therefore, to lower $Rm$ and $\Lambda$. The decrease in $f_{ohm}$ means that Ohmic dissipation becomes less important than viscous dissipation. The magnetic energy ratio, $E_{mag}/E_{kin}$, and the two dipolarities, $f_{dip}$ and $f_{dip,dyn}$, increase with $Pr$ because it leads to a more efficient dynamo mechanism. 

With these trends in mind, we describe the effect of the constant $Pr$ and $Pm$ assumptions on the obtained trends in Figs. \ref{Fig: General diagnostics} and \ref{Fig: Bdyn vs time}. A slight increase in $Pr$ is expected to occur during the long-term evolution of the planets because while it cools down, the ratio of the thermal and electrical conductivities (inversely proportional to $Pr$) decreases according to the Widemann-Franz law, which is valid in the metallic region \citep{Frenchetal2012}. In contrast, the viscosity and conductivity themselves are not thought to vary appreciably with temperature (i.e., in time) \citep{Frenchetal2012, Bonitzetal2024}. Therefore, by using the trend of $\Lambda$ with $Pr$, we expect that for an evolutionary change in $Pr$, the trend $B_{dyn}(t)$ would be slightly steeper, which might agree even better with the \cite{Reiners&Christensen2010} scaling law trend. However, a firm conclusion about this based on a modified setup for diffusivities and $Pr$ evolution is left for future work.

In general, the evolutionary trends shown in Fig. \ref{Fig: General diagnostics} are maintained regardless of $Pr$ and $Pm$. $Rm$ decreases similarly for all set runs. Finally, $E_{mag}/E_{kin}$, $\Lambda$, and $f_{ohm}$ show a different behavior depending on their dipolarity. The two most strongly dipolar solutions ($Pm=1, Pr=2$ and $Pm=2, Pr=2$) consistently show the same behavior as noted above, that is, an increase in $E_{mag}/E_{kin}$ and $f_{ohm}$ in time, and a plateau or mild decrease in $\Lambda$. In contrast, the most multipolar set of Prandtl numbers ($Pm=4, Pr=1$) decreases in $f_{ohm}$ and $\Lambda$ and increases slightly less in $E_{mag}/E_{kin}$. The other sets of runs are consistent with a transition from multipolar to dipolar similar to Fig. \ref{Fig: General diagnostics}.

\subsection{Scaling laws}
\label{Sec: Scaling laws}

\begin{figure}[t]
\centerline{\includegraphics[width=\hsize]{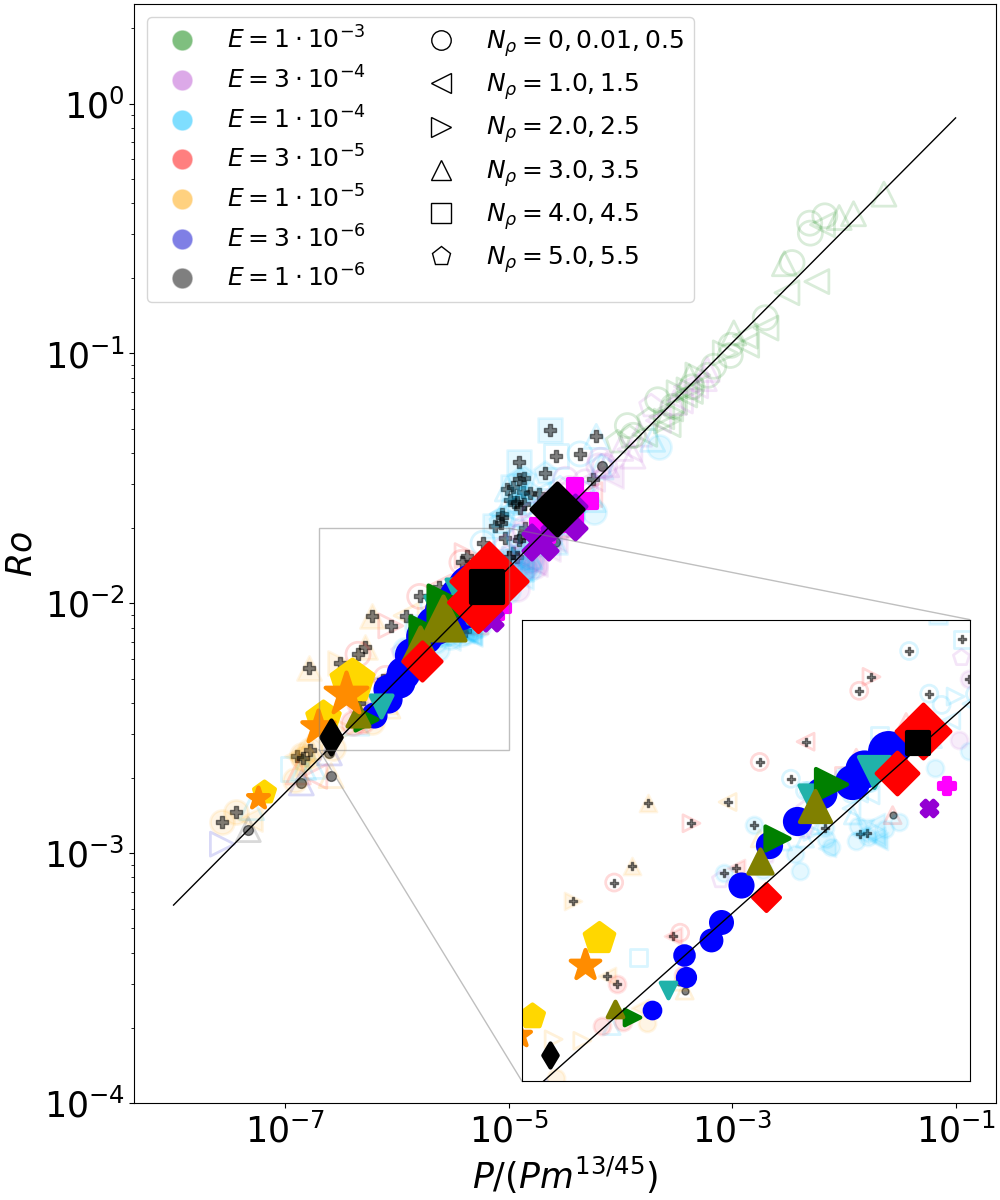}}
\caption{Rossby number as a function of a combination of nondimensional buoyancy power and magnetic Prandtl number. The solid line corresponds to the power law $Ro = 2.47 P^{0.45}Pm^{-0.13}$. The semitransparent data are taken from \cite{Yadavetal2013} and are plotted with a similar symbol and color scheme (the legend is different from our runs). Filled (empty) symbols correspond to dipolar (multipolar) dynamos, and the authors define dipolar as $f_{dip,l<12}^{axi, surf}>0.3$. The Ekman number is color-coded, and the marker shape indicates the degree of density stratification $N_\rho$. Symbols containing a plus have an exponentially decaying conductivity as eq.~(\ref{Eq: Analytical conductivity Gom10}), and those with a dot have a moderate outward decay of $\sigma$, $\nu$, and $\kappa$, all proportional to $\rho(r)$. The complete input and output set can be found in the additional data of the original paper. The results of this work are superposed with the same legend as in Fig. \ref{Fig: General diagnostics}, and the size of the marker denotes the approximate age and mass of the planet.}
\label{Fig: Ro_vs_PPm}
\end{figure}

\begin{figure*}[t]
\centering
\includegraphics[width=.495\textwidth]{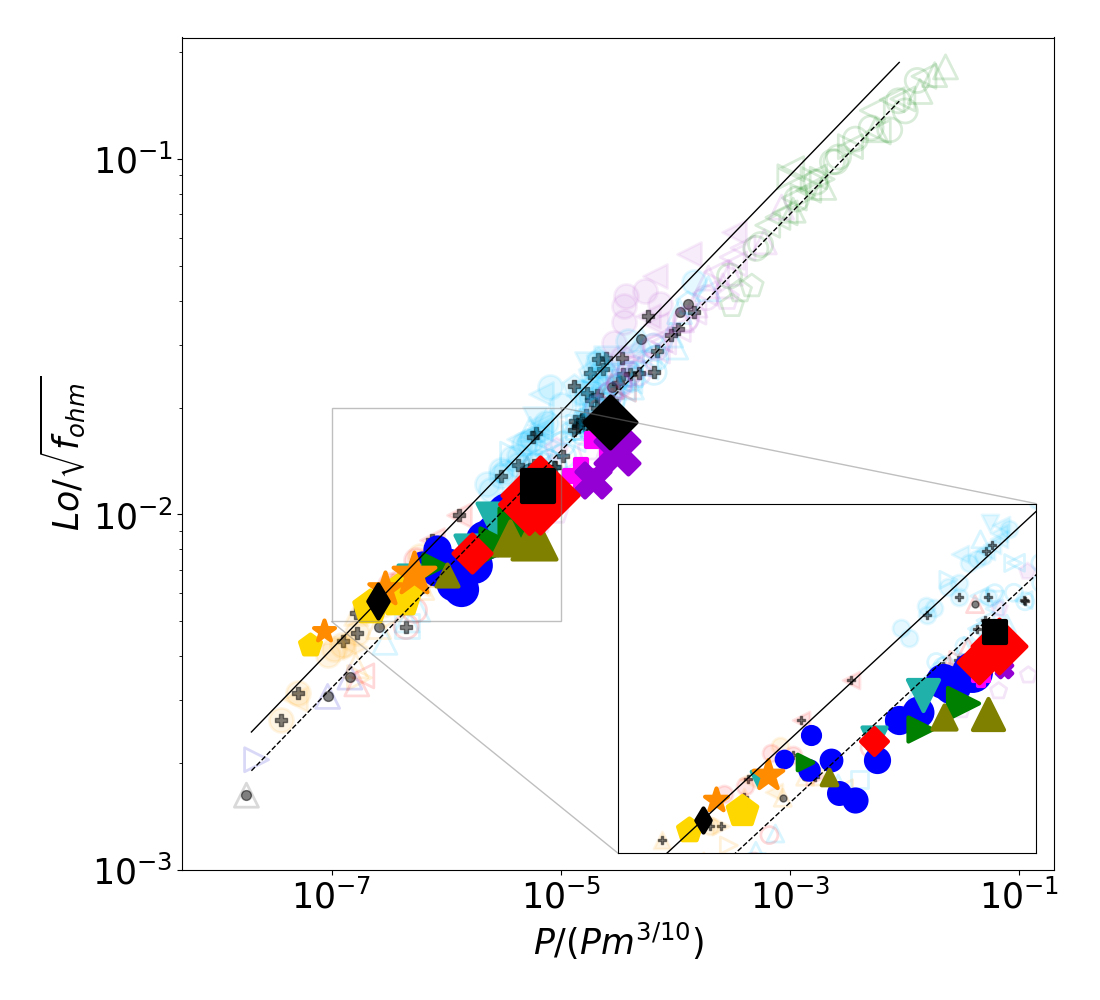}
\includegraphics[width=.46\textwidth]{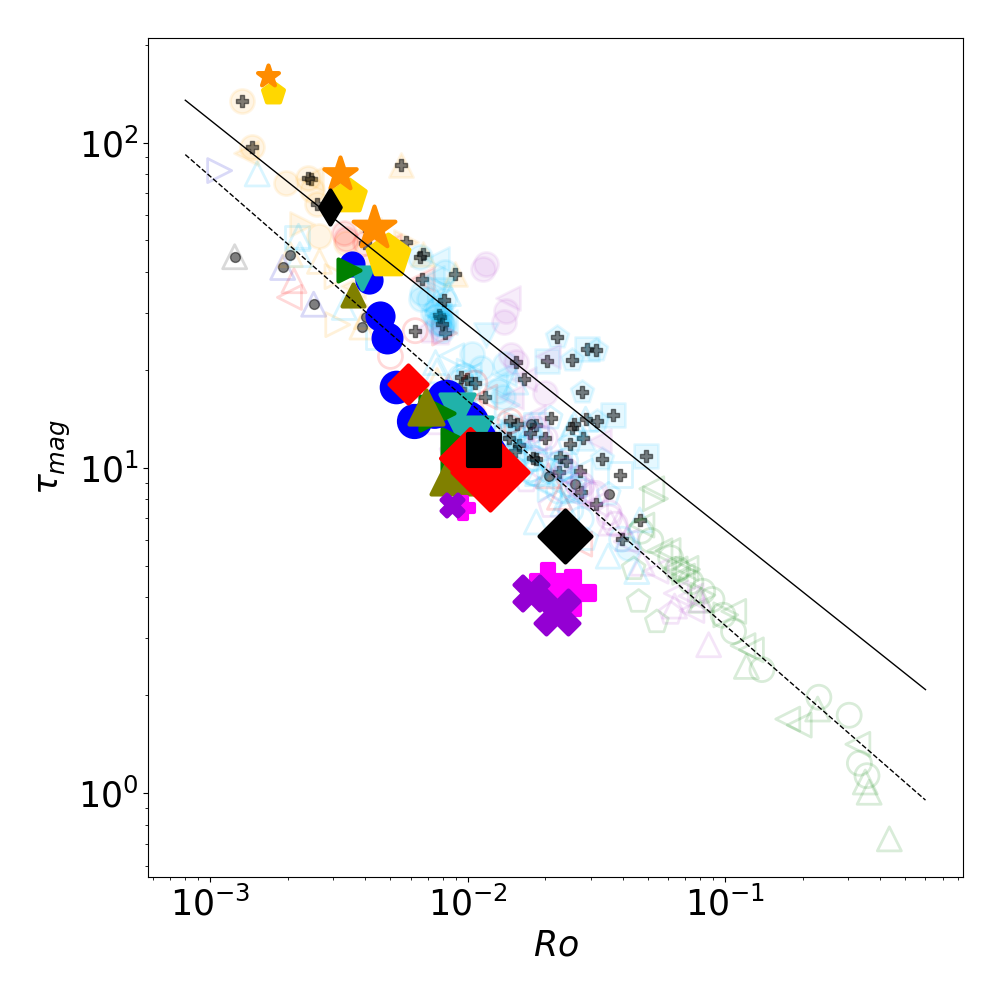}
\caption{Similar to Fig. \ref{Fig: Ro_vs_PPm} (same legend) for magnetically related quantities. On the left: Lorentz number corrected for the fraction of Ohmic dissipation as a function of a combination of nondimensional buoyancy power and magnetic Prandtl number. The scaling relations are $Lo f_{ohm}^{-1/2} = A P^{\frac{1}{3}} Pm^{\frac{1}{10}}$, where $A$ is 0.9 or 0.7 for dipolar and multipolar dynamos, respectively. \cite{Yadavetal2013} found that the value $f_{dip,l<12}^{axi, surf}>0.3$ divides the data into dipolar and multipolar, and these two types of runs are best fit separately. On the right: Characteristic timescale of the magnetic energy dissipation as a function of Rossby number. The scaling relations are $\tau_{mag,dip} = 1.51~Ro^{-0.63}$ and $\tau_{mag,mulip} = 0.67 ~Ro^{-0.69}$.}
\label{Fig: Losqrtfohm_vs_PPm_&_taumag_vs_Ro}
\end{figure*} 

\cite{Yadavetal2013} used a large set of anelastic dynamo numerical solutions to derive scaling laws by relating several representative dimensionless diagnostic parameters. Their dynamos covered a large space of parameters: 0 $\leq N_\rho \lesssim$ 5.5, 0.1 $\leq \eta \leq$ 0.75, 0.3 $\leq Pr \leq$ 10,  0.2 $\leq Pm \leq$ 20,  10$^-{6}$ $\leq E \leq$ 10$^-{3}$, and 2.5 $\cdot$ 10$^5$ $\leq Ra \leq$ 2.5 $\cdot$ 10$^9$. The scattering plots shown in Fig. \ref{Fig: Ro_vs_PPm} and \ref{Fig: Losqrtfohm_vs_PPm_&_taumag_vs_Ro} superimpose the results of \cite{Yadavetal2013} with the data representing our models.

To compare our results, we used the inverse rotation frequency $\Omega^{-1}$ as the time unit and the magnetic field units of $\Omega D \sqrt{\mu_o \rho_o}$. The buoyancy power $P$ is 
\begin{equation}
    P = \frac{Ra E^3}{Pr} \frac{\langle  \tilde{\alpha} \tilde{T} \tilde{g} s' u_r \rangle}{M}~,
    \label{Eq: Power in rotation time units}
\end{equation}
where $M$ is the dimensionless mass of the shell and is listed in Table~\ref{Tab: Cold Jupiter input parameters} for our models. In Fig.~\ref{Fig: Ro_vs_PPm} we show $Ro$ as a function of $P/Pm^{13/45}$. For the runs of \cite{Yadavetal2013}, there is a clear separation between the models with constant $\sigma$, which lie close to the best-fitting power law (gray line), and the runs with a decaying $\sigma$ profile near the surface, which lie slightly above but parallel to the trend. The reason is that for a decaying $\sigma$, the strong jets that appear in the external nonconductive layer tend to increase the total kinetic energy, and thus, $Ro$ for the same amount of available $P$. Our definition of $Ro$ differs by a factor of $1/\tilde{\lambda}$ , which erases the outer jet contribution. Our runs therefore mostly lie above the scaling law itself. Our models represent different evolutionary stages of planetary dynamos that move through this dimensionless space. This progression is parallel to the power law $Ro = 2.47 P^{0.45}Pm^{-0.13}$ and reaches from higher to lower values: For a single model, $Ro$ decreases by about half an order of magnitude, while $P/Pm^{13/45}$ decreases by one order of magnitude (which corresponds to the 0.45 exponent). Physically, this evolution should be positioned orders of magnitude away, but if the scaling law holds, so should this trend.

We also defined the Lorentz number, $Lo$, as the nondimensional magnetic field strength in $\Omega^{-1}$ time units per unit of mass. In this case, $B$ is in units of $\Omega d \sqrt{\rho_o \mu_o}$, and we relate the definition of $Lo$ with $\Lambda$, $E$ and $Pm$,
\begin{equation}
    Lo = \frac{B_{rms}}{\Omega d \sqrt{\rho_o \mu_o}} \frac{1}{\sqrt{M}} = \sqrt{\dfrac{2 E_{mag} E^2}{M}}~.
    \label{Eq: Lorentz number in rotation time units}
\end{equation}
The other magnetically related scaling law involves the characteristic timescale of magnetic energy dissipation $\tau_{mag}$, which is defined as the nondimensional magnetic energy divided by the joule heat dissipation (all in units of $\Omega^{-1}$),
\begin{equation}
    \tau_{mag} = \frac{E_{mag} E^2}{P f_{ohm} M}~.
    \label{Eq: tau_mag in rotation time units}
\end{equation}
In Fig. \ref{Fig: Losqrtfohm_vs_PPm_&_taumag_vs_Ro} we show the the scaling laws for $Lo$ and $\tau_{mag}$ with the same legend as in Fig.~\ref{Fig: Ro_vs_PPm}. \cite{Yadavetal2013} reported that dipolar- and multipolar-dominated solutions take similar but parallel trends. We overplot our runs with the joined dipolar and multipolar branches for the two scaling laws (they show them separately). As suspected from Sect. \ref{Sec: Evolutionary changes}, our series evolves from the multipolar to the dipolar branch in both diagrams, but this is more clearly visible in the $\tau_{mag}$ plot. This is also an argument in favor for this possible multipolar to dipolar transition.

The power-law relations shown above are purely fits obtained from \cite{Yadavetal2013}. The velocity scaling of $Ro \propto P^\alpha$ with $\alpha$ somewhat larger than 0.4 was theoretically justified by force balances by some authors \citep{Aubertetal2001, Davidson2013, Starchenko&Jones2002}, but none derived a $Pm$ dependence. Similarly, $\tau_{mag} \propto Ro^\alpha$ was also discussed, where $\alpha \lesssim -1$ \citep{Christensen&Tilgner2004,Stelzer&Jackson2013} or $\alpha \approx -0.75$ \citep{Davidson2013}. Finally, if the magnetic field is only a function of power, dimensional arguments dictate that it must depend on the cubic root of the power, that is, $Lo \propto P^{1/3}$ \citep{Kunnenetal2010, Christensenetal2009, Davidson2013}.

\section{Conclusions}\label{Sec: Conclusions}

We used radial profiles taken from gas giant evolutionary models to obtain sequences of 3D MHD spherical shell dynamo models. From the public code MESA, we obtained the radial hydrostatic profiles for planets with different masses at different stages of evolution: $0.3 \, M_J \le M_P \le 4 \, M_J$ and 0.2 Gyr $\leq t \leq$ 10 Gyr, respectively. From the evolutionary tracks, we derived the trends for the dynamo parameters. Using the radial profiles as the background state, we solved the resistive MHD equations under the anelastic approximation with the pseudospectral spherical shell code MagIC. We obtained saturated dynamo solutions and interpreted them as different snapshots of planetary dynamos during their long-term evolution.

For our longest set of runs that represents different evolutionary times of a 1 $M_J$ planet, we find a transition from a multipolar- to a dipolar-dominated dynamo regime. Within the dipolar or multipolar branch, a few snapshots are enough to generally assess the behavior that cannot be straightforwardly derived from scaling laws. As the planet evolves and cools, we obtained a steady decrease for $Rm$, $P$, and $Ro$ as well as an increase in the volumetric and surface dipolarities. For multipolar dynamo solutions $\Lambda$, $f_{ohm}$ and $E_{mag}/E_{kin}$ decrease with time, whereas they increase  for dipolar dynamos. These quantities are a proxy for the magnetic field energy, the dynamics of the power dissipation, and the energy ratio. These trends hold for different Prandtl numbers and for the 4 $M_J$ models. Future studies are needed to confirm that these trends are also observed at lower Ekman numbers because the physical nondimensional parameter space is currently computationally inaccessible.

The decay of the magnetic field strength at the dynamo surface (Fig.~\ref{Fig: Bdyn vs time}) is roughly compatible with existing scaling-law estimates. Our models are based on a sequence of realistic backgrounds, and thus, they can be representative of the real long-term dynamo evolution. The trends in mass and age are also expected to hold for the realistic but computationally unfeasible range of nondimensional numbers (which is the usual intrinsic caveat of any dynamo study). We compared our results with the anelastic scaling laws of \cite{Yadavetal2013} between nonlinear combinations of dimensionless diagnostics, and we showed that the long-term evolution of a cold Jupiter dynamo evolves in this parameter space. Consistently with the diagnostics on the morphology described above, some age sequences transition from the multipolar-dominated to the dipolar-dominated family of solutions, which are separated in this parameter space, as shown by \cite{Yadavetal2013}.

Additionally, in the sequences, we considered fixed values of $Pr$ and $Pm$, and the values of $Ra$ and $E$ changed only due to the evolving values of $\Delta T$, $T_o$, and the shell thickness. Taking into account the likely slight decrease in $Pr$ (due to the long-term cooling, which decreases the thermal conductivity), we would expect a slightly steeper decrease in the magnetic field at the dynamo surface. In the presence of non-negligible external torques that would spin down the planet (e.g., tidal frictions with large satellites), $E$ would (slightly) increase, which for the trends seen in our sets would lead to a (slight) enhancement of the slow magnetic decay. We leave this fine-tuned exploration for a follow-up work.

Taking Jupiter and Saturn as prototypes, we expect gas-giant dynamos to be dipolar-dominated at an advanced evolutionary age. However, according to our results, multipolar gas-giant dynamos could exist in the early stages of planetary dynamos, which would then evolve into a dipolar regime. Another possibility is that gas-giant dynamos are already born in a dipole-dominated parameter space region and remain so. These arguments can also be applied to mildly irradiated gas giants and brown dwarfs because they have similar low-$Ro$ dynamos that follow the scaling laws of \cite{Christensenetal2009} because the orbital distance is large enough to prevent tidal interactions or inflation from dominating their energy budget. In contrast, these trends cannot be applied to rapidly rotating main-sequence stars as their evolution is dictated by hydrogen burning and not by a slow cooling.

Finally, the NASA exoplanet archive currently (March 2025) lists more than 650 confirmed gas-giant candidates with $M\text{sin}(i)>0.2 \, M_{J}$ and $P_{orb} >$ 200 days, for which irradiation and tidal synchronization are negligible. Only four exoplanets under 10 pc have an estimate for the Solar System age. The closest and youngest exoplanet ($0.6\pm0.2$ Gyr) is $\epsilon$ Eridani b \citep{Hatzesetal2000}, with a mass of $0.66_{-0.09}^{0.12}$ $M_J$, which is a good candidate for an intense multipolar-dominated dynamo. The other three candidates, Gliese 832 b \citep{Baileyetal2009}, HD 219134 h \citep{Motalebietal2015, Vogtetal2015}, and GJ 3512 b \citep{Moralesetal2019}, with predicted ages $>$5 Gyr, are likely to instead host Jupiter-like dipolar dynamos. If their magnetic field is intense enough to produce electron-cyclotron maser emission that is detectable from ground (with an associated gyro-frequency, $\nu \simeq 2.8~B$[G] MHz, higher than the ionospheric $\approx$10 MHz cutoff; \citealt{Zarka1998}), current (LOFAR) and next-generation (SKA-low) low-frequency radio interferometers might eventually detect their exoplanetary radio emission. This might provide indications of the intensity of the magnetic fields, and, possibly, their morphology.

\begin{acknowledgements}
AEL and CSG's work has been carried out within the framework of the doctoral program in Physics of the Universitat Autònoma de Barcelona. AEL, DV, CSG, and TA are supported by the European Research Council (ERC) under the European Union’s Horizon 2020 research and innovation program (ERC Starting Grant "IMAGINE" No. 948582, PI: DV). FDS acknowledges support from a Marie Curie Action of the European Union (Grant agreement 101030103). AEL, FDS, DV, CSG, and TA acknowledge the support from the ``Mar\'ia de Maeztu'' award to the Institut de Ciències de l'Espai (CEX2020-001058-M). We acknowledge the use of the MareNostrum BSC supercomputer of the Spanish Supercomputing Network, via projects RES/BSC Call AECT-2024-2-0011 (PI AEL), AECT-2023-2-0034 (PI FDS), AECT-2024-2-0003 (PI FDS). AEL acknowledges support and hospitality from the Simons Foundation through the predoctoral program at the Center for Computational Astrophysics, Flatiron Institute. We are grateful to Thomas Gastine, for guidance on the use of MagIC. The authors acknowledge insightful comments by the anonymous referee which helped to signiﬁcantly improve the manuscript.
\end{acknowledgements}

\bibliographystyle{aa}
\bibliography{biblio}

\begin{appendix}

\section{Sensitivity on the density ratio}
\label{App: Sensitivity on the density ratio}

To evaluate how the external cut applied to the 1D MESA profile influences the overall dynamo behavior, we compare the dynamo corresponding to $N_\rho \approx 1.1, 3.0, 3.7, 4.6$ (i.e., $\rho_o/\rho_i \approx 10, 20, 40, 100$) for the 1$M_J$ 1 Gyr model keeping the same 3D resolution of ($N_r$,$N_{\theta}$,$N_{\phi}$) = (289,256,512). By looking at the spectra, the model with $N_\rho \approx $ 4.6 seems slightly under-resolved (there is an overall drop of only 1 order of magnitude, less than what indicates a large enough grid), but the overall quantities seem to follow the same trend as with the other distinct $N_\rho$ models. We could not explore properly higher values of $N_\rho\gtrsim 5.3$, due to the excessive resolution required. In Table \ref{Tab: Different density ratios diagnostics} we show the different diagnostics.

\begin{table}[ht]
\centering
\scriptsize
\caption{General outputs for the 1 $M_J$ 1 Gyr models with different densities.}
\begin{tabular}{@{}cccccccc@{}}
\hline \hline \\[-2.0ex]
  $N_\rho$           & 2.30                 & 2.99                 & 3.68                 & 4.58                  \\ \hline \ \\[-2.0ex]  
  $Rm$               & 1097                 & 732                  & 461                  & 301                   \\   
  $Ro$               & 1.35$\cdot$10$^{-2}$ & 8.20$\cdot$10$^{-3}$ & 4.98$\cdot$10$^{-3}$ & 3.16$\cdot$10$^{-3}$  \\  
  $\Lambda$          & 1.87                 & 0.67                 & 0.172                & 0.072                 \\  
  $P$                & 8.74$\cdot$10$^{10}$ & 7.16$\cdot$10$^{10}$ & 5.10$\cdot$10$^{10}$ & 3.83$\cdot$10$^{10}$  \\  
  $E_{mag}/E_{kin}$  & 0.207                & 0.146                & 0.090                & 0.087                 \\  
  $f_{ohm}$          & 0.283                & 0.190                & 0.101                & 0.066                 \\ 
  $E_{kin}$ (code)   & 1.48$\cdot$10$^7$    & 1.62$\cdot$10$^7$    & 1.54$\cdot$10$^7$    & 1.58$\cdot$10$^7$     \\ 
  $E_{kin}$ (erg)    & 1.39$\cdot$10$^{38}$ & 9.3$\cdot$10$^{37}$  & 4.80$\cdot$10$^{37}$ & 1.82$\cdot$10$^{37}$  \\ 
  $E_{mag}$ (code)   & 3.07$\cdot$10$^6$    & 2.37$\cdot$10$^6$    & 1.38$\cdot$10$^6$    & 1.37$\cdot$10$^5$     \\ 
  $E_{mag}$ (erg)    & 2.90$\cdot$10$^{37}$ & 1.37$\cdot$10$^{37}$ & 4.32$\cdot$10$^{36}$ & 1.58$\cdot$10$^{36}$  \\ 
  $f_P$ (\%)         & 0.18                 & 0.0033               & 0.31                 & 0.43                  \\ \hline \hline \\[-2.0ex] 
\end{tabular}
\label{Tab: Different density ratios diagnostics}
\end{table}

Generally, many of the dimensionless quantities are affected by the value of $N_\rho$. Except for the $N_\rho \approx 100$ model, the overall kinetic energy seems to plateau (in code units). However, since a dominant fraction of the kinetic energy is located in the nonconductive outer layers (see Fig. \ref{Fig: Energy spectra and radial distribution for 1MJ OK models}), all magnitudes containing $u_{rms}$ may differ substantially. As we capture more density ratio, the aforementioned zonal flows gain importance and compete against the magnetic field in the interior affecting the overall dynamics. For example, even though $Ra$ increases, the total buoyant power, $P$, decreases because it depends only on $u_r$. Similarly, $Rm$ and $Ro$ also decrease even though they depend on $u_{rms}$. This is due to the decreasing dimensionless conductivity $1/\tilde{\lambda}$, which erases the zonal flow contribution and only captures the suppression of the internal convection with higher $N_\rho$. Consequently, the Elsasser number $\Lambda$ and $E_{mag}$, also $N_\rho$ also decrease for the same suppression reasons. Correcting factors for the dimensionless mass $M$ and volume $V$ do not mitigate the differences among the $N_\rho$ series. 

To overcome this systematic effect, we tend to compare models with a similar $N_\rho$, independently of the mass and age of the model. For most models, we choose $N_\rho \approx 3$, as it is more computationally feasible but still has a relevant nonconductive outer layer where the zonal jet develops. This restriction allows us to analyze the different saturated models for an evolutionary sequence.

\section{Initial conditions from previous models}
\label{App: Continuing models from past models}

\begin{figure}[t]
  \centering
{\includegraphics[width=.85\hsize]{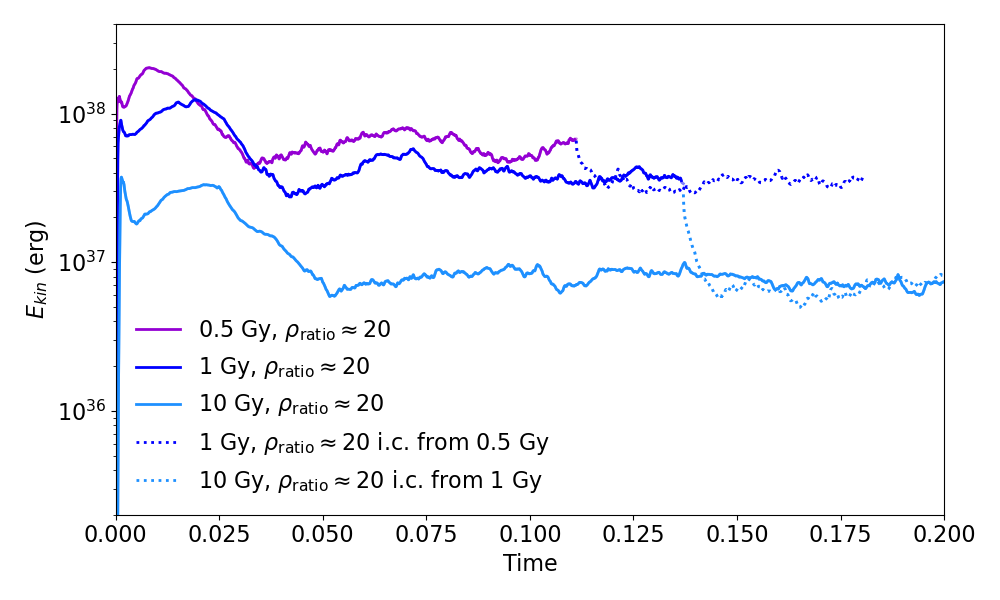} \label{Fig: kinetic energy continuation}} \\{\includegraphics[width=.85\hsize]{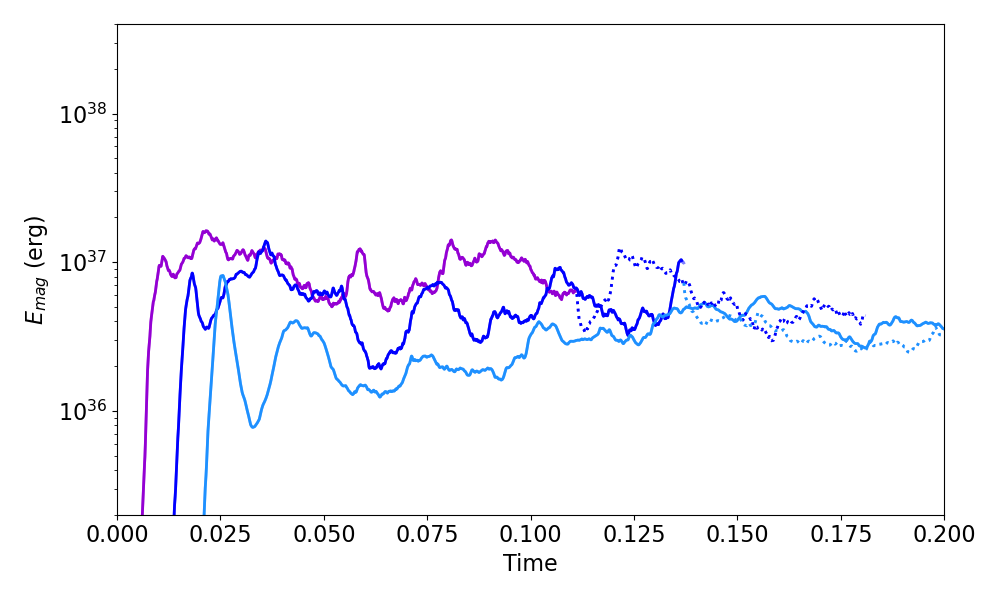} \label{Fig: magnetic energy continuation}}
  \caption{Kinetic (top) and magnetic (bottom) energy evolution time series for the 1 MJ models at 3 different ages (different colors). Solid lines are simulations starting from a ${\mathbf u}=0$ initial conditions, while the dotted lines take as initial condition a snapshot of the saturated solution of another model. Time is in viscous units.}
  \label{Fig: Timeseries for kinetic and magnetic energies continuation}
\end{figure}

To reduce computational resources and avoid starting each model with random initial conditions, we have usually employed already saturated solutions as initial conditions for other models. As we are interpreting the different models as stages in planetary evolution an obvious choice would be to use, for example, the saturated state of the 1 $M_J$ 0.5 Gyr model as initial conditions for the $M_J$ 0.7 Gy, and so on.

As a proof of concept, we made two tests: we used the final saturated state of a 0.5 Gyr model as initial conditions for a 1 Gyr model, and the same for a 1 Gyr and 10 Gyr models. In Fig. \ref{Fig: Timeseries for kinetic and magnetic energies continuation} we show the kinetic and magnetic energy time series for this transitions. The steady states reached are indistinguishable, that is one cannot discern from the spectra, radial distribution, or final diagnostics in which initial conditions were used. The only quantities that showed a noticeable difference are the dipolarity indicators ($f_{dip}$), but they are within one standard deviation from each other. To obtain satisfactorily similar means much longer computing times would be probably required. Overall, there is convergence, starting from different initial conditions.

The activation of convection, followed by the dynamo kinematic phase and finally to the saturated phase where Lorentz forces become relevant is a lengthy computation. Starting from an already saturated solution not too far from the expected one highly reduces by more than a factor of 5 the computing time needed to reach the new steady state. Thus in most of all models, we have used a high $Ra$ model as initial conditions, specifically the saturated 1 $M_J$ 0.5 Gyr model with $Pm = Pr = 1$.

\section{Spectral-radial distributions and dynamo surface definition}
\label{App: Dynamo surface definition}

\begin{figure*}
\centering{
\includegraphics[width=.325\textwidth]{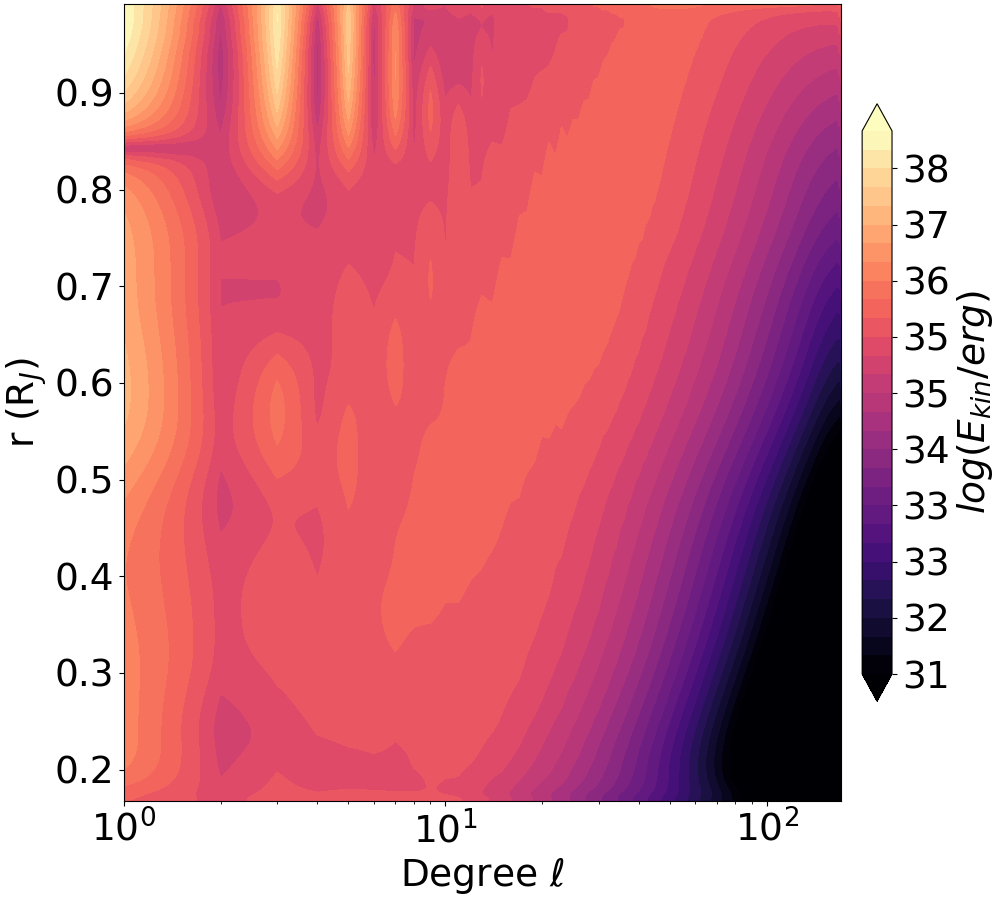}
\includegraphics[width=.325\textwidth]{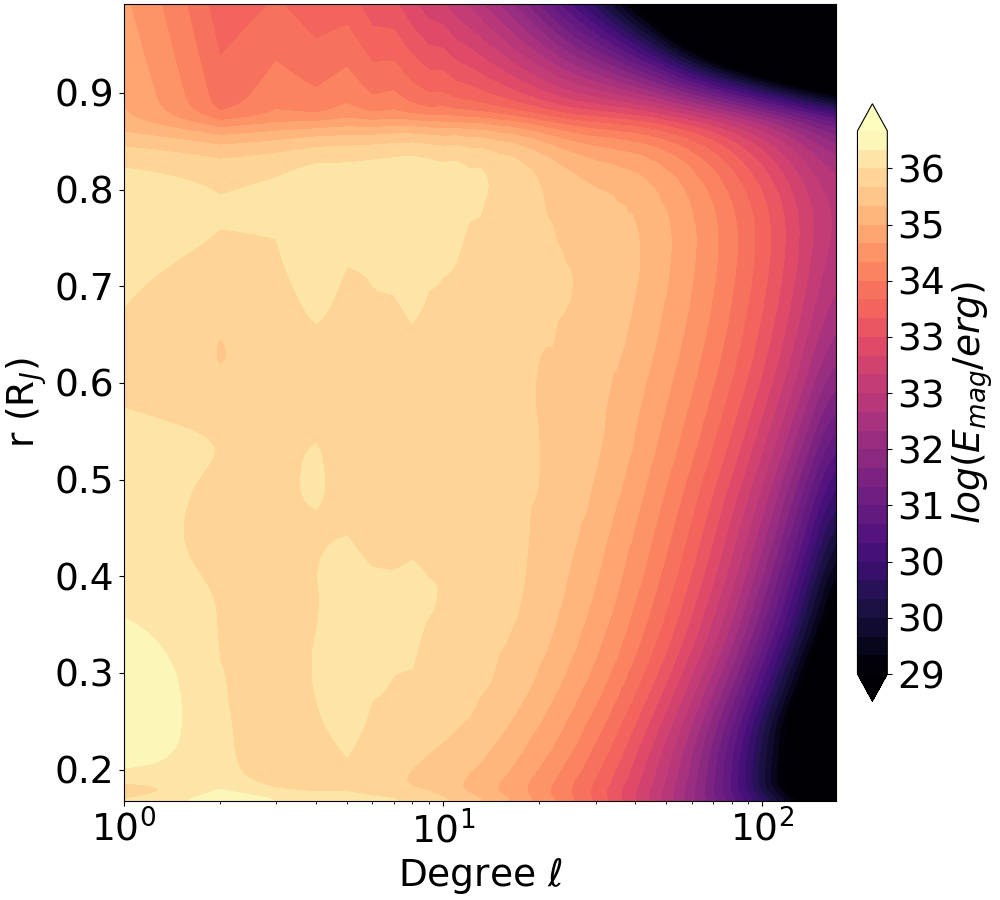}
\includegraphics[width=.325\textwidth]{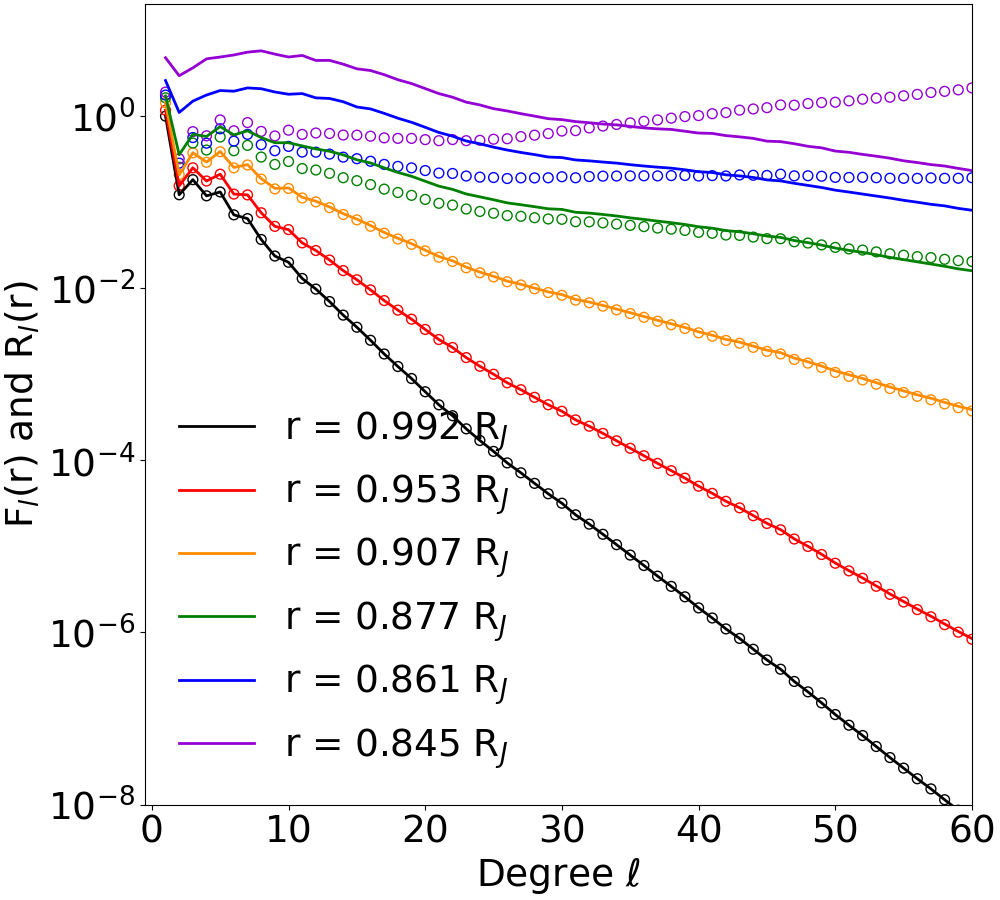}}
\caption{Energy distributions of the of the 1 M$_J$ 2.1 Gyr model. On the left and center: Spectral-radial kinetic and magnetic energy distributions. On the right: Lowes spectra $R_l(r)$ (circles) superimposed with $F_l(r)$ (solid lines) at different depths for the saturated dynamo solution of the same model.}
\label{Fig: 2D spectra and Lowes spectra}
\end{figure*} 

\begin{figure}[t]
  \centering
  \includegraphics[width=.8\hsize]{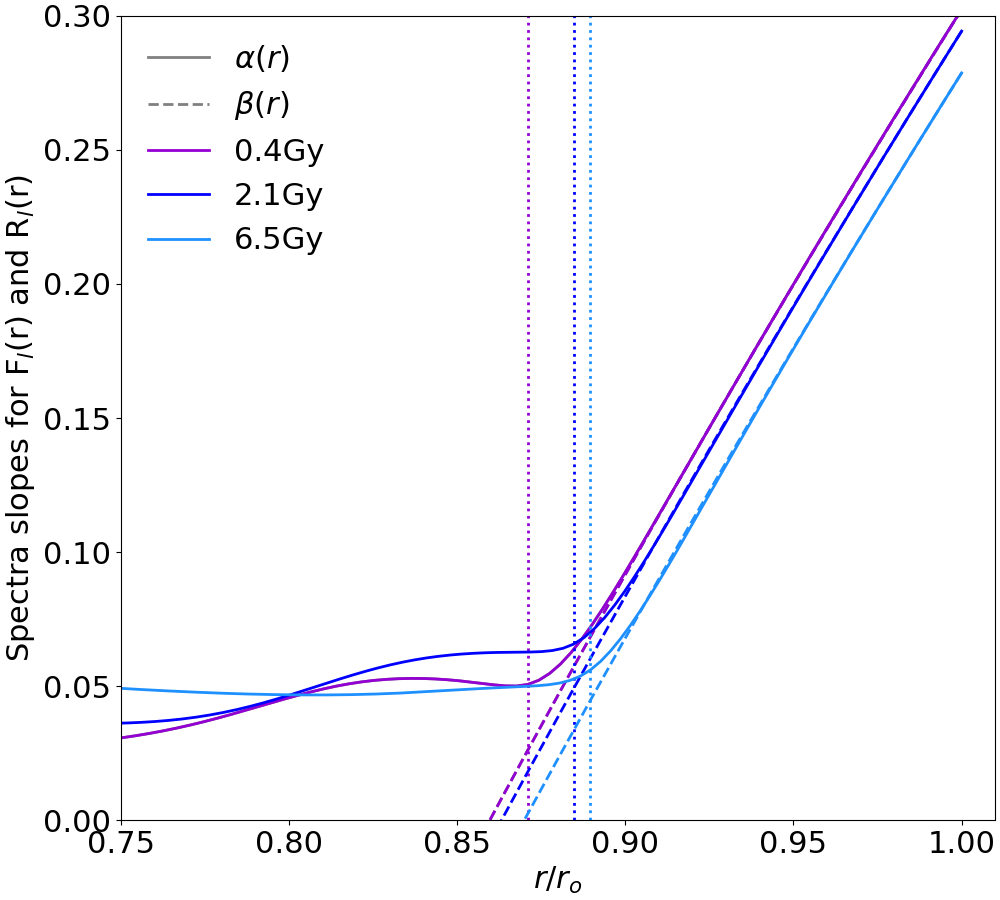}
  \caption{Spectral slopes at different dimensionless radial depths for the same runs as Fig. \ref{Fig: Energy spectra and radial distribution for 1MJ OK models}. The vertical lines are their respective $r_m$.}
  \label{Fig: Spectra slopes}
\end{figure}

To compare our results with \citep{Tsang&Jones2020}, we obtained the spectral-radial energy distribution. In the left and middle panels of Fig.~\ref{Fig: 2D spectra and Lowes spectra} we show these spectra for the saturated 1 $M_J$ 2.1 Gyr model. The radial integration of both spectra is already shown in Fig.~\ref{Fig: Energy spectra and radial distribution for 1MJ OK models}. It can be observed that the radially dependent magnetic energy spectra, $F_l(r)$, decay rapidly for $r>r_m=0.88~R_J$. At the same region, the kinetic spectra start showing the equatorial jet pattern.

In the region where $\mathbf{J}$ = 0, we define the scalar potential $V$, for which $\mathbf{B} = - \boldsymbol{\nabla} V$, with its usual spherical expansion:
\begin{equation}
	V \equiv a \sum_{l=1}^{l_{max}} \left( \frac{a}{r} \right)^{l+1} \sum_{m=0}^{l} P_l^m(cos\theta) \left[ g_l^mcos(m\phi)+h_l^msin(m\phi)  \right]~,
	\label{Internal potential spherical harmonics}
\end{equation}
where $a$ is taken as the planetary radius, $P_l^m$ are the Schmidt semi-normalized associated Legendre polynomials, and the Gauss coefficients $g_l^m$ and $h_l^m$ are obtained from measurements. Then the Lowes spectrum \citep{Mauersberger1956,Lowes74} on the planetary surface is defined as:
\begin{equation}
	R_l(a) \equiv (l+1)\sum_{m=0}^l \left[ (g_l^m)^2 + (h_l^m)^2 \right]~.
\end{equation}
For any other radii, each degree $l$ has a different contribution as they decay differently with distance $r$:
\begin{equation}
	R_l(r) = \left(\frac{a}{r}\right)^{2l+4} R_l(a)~.
\end{equation}
This expression gives the Lowes spectrum in the interior of the planet, that is, the downward extrapolation which would be valid in the absence of electrical current (potential field). Therefore, we expect to have Lowes spectra only outside the dynamo surface, since inside it $\mathbf{B}$ stops being potential. Moreover, from equipartition reasons, it is usually assumed that the magnetic field in the dynamo region is equally distributing among different scales until the diffusive scale ('white source hypothesis’, \citealt{Backus1996}). A flat spectrum at the dynamo surface would imply that at the planetary surface, there will be a linear relation $log_{10}R_l(r) \approx -\beta(r)$. For our saturated dynamo solutions, we define the Lowes spectrum as the magnetic spectra at the outermost radius, where we impose potential boundary conditions:
\begin{equation*}
    R_l(r_{o}) = F_l(r_{o})~.
\end{equation*}
On the right panel of Fig.~\ref{Fig: 2D spectra and Lowes spectra} we show both $R_l(r)$ and $F_l(r)$ at different radii for one specific saturated solution. The black line corresponds to $r_o$ and the others are at different depths. At some specific depth $R_l(r)$ stops being similar to $F_l(r)$ and even has an unphysical negative slope. The Lowes radius is defined where $\beta(r)=0$ and it is usually taken as the depth where the dynamo starts. As \citep{Tsang&Jones2020} noted, $R_l(r)$ is already quite different from $F_l(r)$ at such radius. To surpass this discrepancy, they similarly defined the spectra slope for $F_l(r)$, that is, $log_{10}F_l(r) \approx -\alpha(r)$. The slope $\alpha(r)$ is almost the same as $\beta(r)$ outside the dynamo region but becomes more or less flat at positive values inside. They argue that radius where $\alpha(r)$ and $\beta(r)$ show a discrepancy is where the effective dynamo surface is.

We similarly obtained the spectra slopes $\alpha(r)$ and $\beta(r)$ by a least square minimization between multipoles 10 and 50, which is the region where spectra are exponential and have not reached the dissipation scales set by our resolution. In Fig.~\ref{Fig: Spectra slopes} we show the resulting spectra slopes, for the same models shown in Fig. \ref{Fig: Energy spectra and radial distribution for 1MJ OK models}. For some of our models, we do not obtain a flat spectrum inside, but if we take the average value of $\alpha(r)$ in the dynamo region and interpolate it for the decaying outer part, we obtain values very similar to the ones of $r_m$, shown as vertical lines. Therefore, we can safely assume that the radial position where the conductivity starts the exponential decay, $r_m$, is a good definition as the dynamo surface for our models.

\section{Diagnostics for all the models}
\label{App: Tabulated runs}

In the Table \ref{Tab: Output data table} we show the output parameters graphically shown in Figs. \ref{Fig: General diagnostics}, \ref{Fig: General diagnostics Pm}.

\begin{table*}[h!]
\centering
\scriptsize
\caption{Output quantities for the dynamo saturated states of all models of Table \ref{Tab: Cold Jupiter input parameters}.}
\begin{tabular}{@{}ccccccccccccccc@{}}
\hline \hline \\[-2.0ex]
  Model          & $N_\rho$ & E   & Ra    &  Pm   &  Pr    &  Rm    &$\Lambda$ &  E$_{mag}$ & E$_{kin}$ &$f_{dip}$ &$f_{ohm}$ & $P_{\nu}$ &  $f_P(\%)$ \\  \hline \\[-2.0ex]
  1$M_J$ 0.2 Gyr & 2.99   & 1.01$\cdot$10$^{-5}$ & 1.63$\cdot$10$^{9}$ &  1 &  1 &1107&1.38  &5.42$\cdot$10$^{6}$&3.44$\cdot$10$^{7}$ &0.0271 &0.235 & 2.123$\cdot$10$^{11}$& 2.4 \\
  1$M_J$ 0.4 Gyr & 2.99   & 1.08$\cdot$10$^{-5}$ & 1.27$\cdot$10$^{9}$ &  1 &  1 &956 &0.99  &3.8$\cdot$10$^{6}$ &3.13$\cdot$10$^{7}$ &0.028  &0.197 &1.49$\cdot$10$^{11}$ & 0.45  \\
  1$M_J$ 0.5 Gyr & 2.98   & 1.08$\cdot$10$^{-5}$ & 1.22$\cdot$10$^{9}$ &  1 &  1 &904 &0.987 &3.7$\cdot$10$^{6}$ &2.56$\cdot$10$^{7}$ &0.031  &0.194 &1.29$\cdot$10$^{11}$ & 0.22  \\
  1$M_J$ 0.7 Gyr & 2.99   & 1.12$\cdot$10$^{-5}$ & 1.05$\cdot$10$^{9}$ &  1 &  1 &829 &0.71  &2.58$\cdot$10$^{6}$&2.54$\cdot$10$^{7}$ &0.028  &0.176 &9.27$\cdot$10$^{10}$ & 0.31  \\
  1$M_J$ 1 Gyr   & 2.99   & 1.12$\cdot$10$^{-5}$ & 9.81$\cdot$10$^{8}$ &  1 &  1 &734 &0.649 &2.31$\cdot$10$^{6}$&1.68$\cdot$10$^{7}$ &0.030  &0.172 &7.36$\cdot$10$^{10}$ & 0.025 \\
  1$M_J$ 1.5 Gyr & 2.98   & 1.15$\cdot$10$^{-5}$ & 8.78$\cdot$10$^{8}$ &  1 &  1 &641 &0.385 &1.32$\cdot$10$^{6}$&1.22$\cdot$10$^{7}$ &0.0267 &0.141 &5.42$\cdot$10$^{10}$ & 0.062 \\
  1$M_J$ 2.1 Gyr & 2.99   & 1.18$\cdot$10$^{-5}$ & 7.87$\cdot$10$^{8}$ &  1 &  1 &525 &0.44  &1.29$\cdot$10$^{6}$&8.61$\cdot$10$^{6}$ &0.045  &0.191 &4.10$\cdot$10$^{10}$ & 0.066 \\
  1$M_J$ 2.8 Gyr & 2.99   & 1.20$\cdot$10$^{-5}$ & 7.33$\cdot$10$^{8}$ &  1 &  1 &439 &0.490 &1.52$\cdot$10$^{6}$&7.1$\cdot$10$^{6}$  &0.104  &0.216 &3.32$\cdot$10$^{10}$ & 0.18  \\
  1$M_J$ 3.5 Gyr & 2.99   & 1.21$\cdot$10$^{-5}$ & 6.71$\cdot$10$^{8}$ &  1 &  1 &400 &0.39  &1.85$\cdot$10$^{6}$&6.16$\cdot$10$^{6}$ &0.114  &0.207 &2.93$\cdot$10$^{10}$ & 0.26  \\
  1$M_J$ 5 Gyr   & 2.99   & 1.23$\cdot$10$^{-5}$ & 6.22$\cdot$10$^{8}$ &  1 &  1 &370 &0.51  &1.94$\cdot$10$^{6}$&4.89$\cdot$10$^{6}$ &0.130  &0.232 &2.32$\cdot$10$^{10}$ & 0.20  \\
  1$M_J$ 6.5 Gyr & 2.98   & 1.28$\cdot$10$^{-5}$ & 5.52$\cdot$10$^{8}$ &  1 &  1 &323 &0.53  &2.57$\cdot$10$^{6}$&4.40$\cdot$10$^{6}$ &0.114  &0.247 &2.15$\cdot$10$^{10}$ & 0.079 \\
  1$M_J$ 10 Gyr  & 3.00   & 1.30$\cdot$10$^{-5}$ & 4.97$\cdot$10$^{8}$ &  1 &  1 &274 &0.558 &2.08$\cdot$10$^{6}$&3.65$\cdot$10$^{6}$ &0.130  &0.250 &1.51$\cdot$10$^{10}$ & 0.043 \\ \hline \\[-2.0ex]
  1$M_J$ 1   Gyr & 2.30   & 1.23$\cdot$10$^{-5}$ & 7.92$\cdot$10$^{8}$ &  1 &  1 &1092&1.87  &3.07$\cdot$10$^{6}$&1.48$\cdot$10$^{7}$ &0.0057 &0.283 &8.74$\cdot$10$^{10}$ & 0.18  \\
                 & 3.68   & 1.08$\cdot$10$^{-5}$ & 1.10$\cdot$10$^{9}$ &  1 &  1 &461 &0.172 &1.38$\cdot$10$^{6}$&1.54$\cdot$10$^{7}$ &0.053  &0.101 &5.10$\cdot$10$^{10}$ & 0.31  \\
                 & 4.58   & 1.05$\cdot$10$^{-5}$ & 1.20$\cdot$10$^{9}$ &  1 &  1 &301 &0.072 &1.37$\cdot$10$^{6}$&1.58$\cdot$10$^{7}$ &0.041  &0.066 &3.83$\cdot$10$^{10}$ & 0.43  \\  \hline \\[-2.0ex]
  0.3$M_J$ 5 Gyr & 1.10    & 3.61$\cdot$10$^{-5}$ & 3.18$\cdot$10$^{7}$ &  1 &  1 &213 &0.48  &1.49$\cdot$10$^{5}$&8.3$\cdot$10$^{5}$  &0.030  &0.28  &9.9$\cdot$10$^{8}$   & 4.4  \\
  0.7$M_J$ 5 Gyr & 2.98   & 1.37$\cdot$10$^{-5}$ & 4.01$\cdot$10$^{8}$ &  1 &  1 &213 &0.300 &1.41$\cdot$10$^{6}$&2.55$\cdot$10$^{6}$ &0.131  &0.305 &5.31$\cdot$10$^{9}$  & 0.033\\
  2$M_J$ 5 Gyr   & 2.98   & 1.12$\cdot$10$^{-5}$ & 1.24$\cdot$10$^{9}$ &  1 &  1 &1032&3.14  &1.03$\cdot$10$^{7}$&1.82$\cdot$10$^{7}$ &0.055  &0.358 &2.26$\cdot$10$^{11}$ & 0.90 \\
  4$M_J$ 5 Gyr   & 2.98   & 1.08$\cdot$10$^{-5}$ & 2.49$\cdot$10$^{9}$ &  1 &  1 &2203&14.0  &3.57$\cdot$10$^{7}$&5.98$\cdot$10$^{7}$ &0.0041 &0.526 &1.020$\cdot$10$^{12}$& 15   \\ \hline \\[-2.0ex]
  1$M_J$ 0.5 Gyr & 2.98   & 1.08$\cdot$10$^{-5}$ & 1.22$\cdot$10$^{9}$ & 0.5&  1 &514 &0.41  &3.0$\cdot$10$^{6}$ &3.2$\cdot$10$^{7}$  &0.076  &0.228 &1.335$\cdot$10$^{11}$& 0.16  \\  
                 & 2.98   & 1.08$\cdot$10$^{-5}$ & 1.22$\cdot$10$^{9}$ &  2 &  1 &1790&1.93  &3.50$\cdot$10$^{6}$&2.47$\cdot$10$^{7}$ &0.0045 &0.218 &1.32$\cdot$10$^{11}$ & 5.4   \\
                 & 2.98   & 1.08$\cdot$10$^{-5}$ & 1.22$\cdot$10$^{9}$ &  4 &  1 &3250&5.30  &4.86$\cdot$10$^{6}$&2.27$\cdot$10$^{7}$ &0.00204&0.331 &1.40$\cdot$10$^{11}$ & 53    \\
                 & 2.98   & 1.08$\cdot$10$^{-5}$ & 1.22$\cdot$10$^{9}$ & 0.5& 0.5&1190&2.11  &1.50$\cdot$10$^{7}$&8.10$\cdot$10$^{7}$ &0.020  &0.292 &5.78$\cdot$10$^{11}$ &  6    \\
                 & 2.98   & 1.08$\cdot$10$^{-5}$ & 1.22$\cdot$10$^{9}$ &  1 & 0.5&2050&4.11  &1.46$\cdot$10$^{7}$&7.16$\cdot$10$^{7}$ &0.00934 &0.332 &5.66$\cdot$10$^{11}$ &  11   \\
                 & 2.98   & 1.08$\cdot$10$^{-5}$ & 1.22$\cdot$10$^{9}$ &  1 &  2 &454 &0.213 &8.7$\cdot$10$^{5}$ &7.62$\cdot$10$^{6}$ &0.073  &0.127 &2.84$\cdot$10$^{10}$ & 0.086 \\
                 & 2.98   & 1.08$\cdot$10$^{-5}$ & 1.22$\cdot$10$^{9}$ &  2 &  2 &803 &0.803 &1.63$\cdot$10$^{6}$&6.12$\cdot$10$^{6}$ &0.054  &0.180 &3.07$\cdot$10$^{10}$ & 0.22  \\ \hline \\[-2.0ex]
  1$M_J$ 1 Gyr   & 2.99   & 1.12$\cdot$10$^{-5}$ & 9.81$\cdot$10$^{8}$ & 0.5&  1 &440 &0.25  &1.81$\cdot$10$^{6}$&2.4$\cdot$10$^{7}$  &0.093  &0.153 &7.57$\cdot$10$^{10}$ & 0.061 \\
                 & 2.99   & 1.12$\cdot$10$^{-5}$ & 9.81$\cdot$10$^{8}$ &  2 &  1 &1360&1.31  &2.33$\cdot$10$^{6}$&1.45$\cdot$10$^{7}$ &0.0088 &0.186 &7.57$\cdot$10$^{10}$ & 1.36  \\ 
                 & 2.99   & 1.12$\cdot$10$^{-5}$ & 9.81$\cdot$10$^{8}$ &  4 &  1 &2460&3.52  &3.17$\cdot$10$^{6}$&1.25$\cdot$10$^{7}$ &0.00236&0.230 &7.98$\cdot$10$^{10}$ & 20.9  \\
                 & 2.99   & 1.12$\cdot$10$^{-5}$ & 9.81$\cdot$10$^{8}$ & 0.5& 0.5&918 &0.95  &6.6$\cdot$10$^{6}$ &8.7$\cdot$10$^{7}$  &0.0141 &0.213 &3.08$\cdot$10$^{11}$ & 0.65  \\
                 & 2.99   & 1.12$\cdot$10$^{-5}$ & 9.81$\cdot$10$^{8}$ &  1 & 0.5&164 &2.07  &7.2$\cdot$10$^{6}$ &7.1$\cdot$10$^{7}$  &0.0081 &0.245 &3.18$\cdot$10$^{11}$ & 1.54  \\
                 & 2.99   & 1.12$\cdot$10$^{-5}$ & 9.81$\cdot$10$^{8}$ &  1 &  2 &310 &0.221 &9.1$\cdot$10$^{5}$ &3.83$\cdot$10$^{6}$ &0.124  &0.162 &1.461$\cdot$10$^{10}$& 0.037 \\
                 & 2.99   & 1.12$\cdot$10$^{-5}$ & 9.81$\cdot$10$^{8}$ &  2 &  2 &570 &0.73  &1.45$\cdot$10$^{6}$&3.13$\cdot$10$^{6}$ &0.073  &0.203 &1.593$\cdot$10$^{10}$& 0.084 \\ \hline \\[-2.0ex]
  1$M_J$ 10 Gyr  & 3.00   & 1.30$\cdot$10$^{-5}$ & 4.97$\cdot$10$^{8}$ & 0.5&  1 &157 &0.252 &2.01$\cdot$10$^{6}$&5.0$\cdot$10$^{6}$  &0.333  &0.219 &1.89$\cdot$10$^{10}$ & 0.024 \\
                 & 3.00   & 1.30$\cdot$10$^{-5}$ & 4.97$\cdot$10$^{8}$ &  2 &  1 &532 &1.11  &2.04$\cdot$10$^{6}$&3.30$\cdot$10$^{6}$ &0.062  &0.250 &1.55$\cdot$10$^{10}$ & 0.070 \\ 
                 & 3.00   & 1.30$\cdot$10$^{-5}$ & 4.97$\cdot$10$^{8}$ &  4 &  1 &1100&1.77  &1.59$\cdot$10$^{6}$&3.18$\cdot$10$^{6}$ &0.0139 &0.220 &1.63$\cdot$10$^{10}$ & 3.87  \\
                 & 3.00   & 1.30$\cdot$10$^{-5}$ & 4.97$\cdot$10$^{8}$ & 0.5& 0.5&368 &0.634 &4.29$\cdot$10$^{6}$&1.87$\cdot$10$^{7}$ &0.088  &0.271 &8.113$\cdot$10$^{10}$& 0.23  \\
                 & 3.00   & 1.30$\cdot$10$^{-5}$ & 4.97$\cdot$10$^{8}$ &  1 & 0.5&699 &1.40  &4.78$\cdot$10$^{6}$&1.71$\cdot$10$^{7}$ &0.0456 &0.281 &8.49$\cdot$10$^{10}$ & 0.072 \\
                 & 3.00   & 1.30$\cdot$10$^{-5}$ & 4.97$\cdot$10$^{8}$ &  1 &  2 &135 &0.128 &5.13$\cdot$10$^{5}$&7.78$\cdot$10$^{5}$ &0.301  &0.177 &3.14$\cdot$10$^{9}$  & 0.0068\\
                 & 3.00   & 1.30$\cdot$10$^{-5}$ & 4.97$\cdot$10$^{8}$ &  2 &  2 &257 &0.473 &8.8$\cdot$10$^{5}$ &6.76$\cdot$10$^{5}$ &0.085  &0.251 &3.36$\cdot$10$^{9}$  & 0.069 \\ \hline \\[-2.0ex]
  4$M_J$ 0.5 Gyr & 4.60   & 8.37$\cdot$10$^-6$   & 6.50$\cdot$10$^{9}$ &  1 &  1 &1460&4.4   &7.28$\cdot$10$^{7}$&1.250$\cdot$10$^{8}$&0.0058 &0.374 &2.39$\cdot$10$^{12}$ & 3.6  \\ 
  4$M_J$ 1 Gyr   & 4.58   & 8.98$\cdot$10$^{-5}$ & 5.02$\cdot$10$^{9}$ &  1 &  1 &1130&3.3   &5.26$\cdot$10$^{7}$&8.47$\cdot$10$^{7}$ &0.00768&0.353 &1.542$\cdot$10$^{12}$& 0.63 \\ 
  4$M_J$ 10 Gyr  & 4.60   & 1.06$\cdot$10$^{-5}$ & 2.42$\cdot$10$^{9}$ &  1 &  1 &554 &1.1   &1.61$\cdot$10$^{7}$&2.48$\cdot$10$^{7}$ &0.0051 &0.256 &3.28$\cdot$10$^{11}$ & 0.054\\  \hline  \hline \\[-2.0ex]
\end{tabular}
\label{Tab: Output data table}
\tablefoot{Specifically we show, for all of them, the input nondimensional numbers ($E$ and $Ra$ fixed as in Table \ref{Tab: Cold Jupiter input parameters}, and the Prandtl numbers $Pm$, $Pr$), and the diagnostics: $\Lambda$, $E_{mag}$ (in code units), $E_{kin}$ (in code units), dipolarity $f_{dip}$, Ohmic fraction $f_{ohm}$, buoyancy power $P_\nu$, power imbalance $f_P$. The values are time-averaged. The typical standard deviations, here omitted for the sake of space, are shown as error bars in Figs.~\ref{Fig: General diagnostics}, \ref{Fig: General diagnostics Pm}.}
\end{table*}

\end{appendix}

\end{document}